\documentclass[aps,prx,reprint,superscriptaddress]{revtex4-2}

\usepackage{amsmath}
\usepackage{amssymb}
\usepackage{svg}
\usepackage[separate-uncertainty=true]{siunitx}
\definecolor{bluepoli}{RGB}{0,36,179}
\definecolor{redpoli}{RGB}{204,0,51}
\definecolor{greenpoli}{RGB}{45,137,0}
\definecolor{purplepoli}{RGB}{153,102,204}
\definecolor{azzurropoli}{RGB}{51,53,204}
\definecolor{orangepoli}{RGB}{255,124,17}
\usepackage{xcolor}
\usepackage[normalem]{ulem}
\newcommand\mkst{\bgroup\markoverwith
{\textcolor{red}{\rule[.5ex]{2pt}{0.4pt}}}\ULon}
\newcommand\cwst{\bgroup\markoverwith
{\textcolor{blue}{\rule[.5ex]{2pt}{0.4pt}}}\ULon}

\usepackage{hyperref}
\usepackage{physics}
\usepackage{upgreek} 
\usepackage[T1]{fontenc}
\usepackage{booktabs}
\usepackage{float}

\hypersetup{%
	colorlinks,
	linkcolor={bluepoli}, %
	citecolor={redpoli}, %
	urlcolor={redpoli} %
}

\def\QDevaffil{Center for Quantum Devices, Niels Bohr Institute, University of Copenhagen, 2100 Copenhagen, Denmark}
\def\NTNUaffil{Center for Quantum Spintronics, Department of Physics, Norwegian University of Science and Technology, NO-7491 Trondheim, Norway}
\def\NNFaffil{NNF Quantum Computing Programme, Niels Bohr Institute, University of Copenhagen, 2100 Copenhagen, Denmark}
\def\Chalmersaffil{Department of Microtechnology and Nanoscience, Chalmers University of Technology, SE-412 96 Gothenburg, Sweden}
\def\DecayRate{\mathit{\Gamma}_1}
\def\GammaFunction{\Gamma}
\begin{document}
	
	
 \title{Real-time adaptive tracking of fluctuating relaxation rates\\ in superconducting qubits}

	
\author{Fabrizio~Berritta}
\altaffiliation{Current address: Research Laboratory of Electronics, Massachusetts Institute of Technology, Cambridge, MA 02139, USA} 
\email{fabrizio.berritta@mit.edu}
\affiliation{\QDevaffil}
\affiliation{\NNFaffil}
 \author{Jacob~Benestad}
 \affiliation{\NTNUaffil}
 \author{Jan~A.~Krzywda}
 \affiliation{Lorentz Institute for Theoretical Physics \& Leiden Institute of Advanced Computer Science, Universiteit Leiden, 2311 EZ Leiden, The Netherlands}
\author{Oswin~Krause}
	\affiliation{Department of Computer Science, University of Copenhagen, 2100 Copenhagen, Denmark}
 \author{Malthe~A.~Marciniak}
 \affiliation{\QDevaffil}	
 \affiliation{\NNFaffil}		
 \author{Svend~Kr{\o}jer}
 \affiliation{\QDevaffil}	
 \affiliation{\NNFaffil}	
 \author{Christopher~W.~Warren}
 \affiliation{\Chalmersaffil}
 \affiliation{\QDevaffil}	
 \affiliation{\NNFaffil}		
 \author{Emil~Hogedal}
 \affiliation{\Chalmersaffil}
 \author{Andreas~Nylander}
 \affiliation{\Chalmersaffil} 
 \author{Irshad~Ahmad}
 \affiliation{\Chalmersaffil} 
 \author{Amr~Osman}
 \affiliation{\Chalmersaffil}\author{Janka~Bizn\'{a}rov\'{a}}
 \affiliation{\Chalmersaffil}
 \author{Marcus~Rommel}
 \affiliation{\Chalmersaffil}
 \author{Anita~Fadavi~Roudsari}
 \affiliation{\Chalmersaffil}
 \author{Jonas~Bylander}
 \affiliation{\Chalmersaffil}
 \author{Giovanna~Tancredi}
 \affiliation{\Chalmersaffil} 
 \author{Jeroen~Danon}	\affiliation{\NTNUaffil}
 \author{Jacob~Hastrup}
 \affiliation{\QDevaffil}	
 \affiliation{\NNFaffil}	
 \author{Ferdinand~Kuemmeth}
 \affiliation{\QDevaffil}	
 \affiliation{Institute of Experimental and Applied Physics, University of Regensburg, 93040 Regensburg, Germany}
 \affiliation{QDevil, Quantum Machines, 2750 Ballerup, Denmark}	
 \author{Morten~Kjaergaard}
 \email{mkjaergaard@nbi.ku.dk}
 \affiliation{\QDevaffil}	
 \affiliation{\NNFaffil}	

	
	\date{February 13, 2026}
	\begin{abstract}
    
The fidelity of operations on a solid-state quantum processor is fundamentally bounded by environmental decoherence. Characterizing environmental fluctuations is challenging because the acquisition time of nonadaptive experimental protocols limits temporal precision and can average out rapid features of the underlying dynamics.
Here, we overcome this temporal-resolution limit by two orders of magnitude using a field-programmable gate-array (FPGA) powered classical controller that adaptively and continuously tracks the relaxation-time fluctuations of two fixed-frequency superconducting transmon qubits, which exhibit average relaxation times of approximately $\SI{0.17}{\milli\second}$ and occasionally exceed $\SI{0.5}{\milli\second}$.
We report events in which the relaxation time switches by nearly an order of magnitude over timescales of just tens of milliseconds, rather than minutes or hours as previously reported.
Our real-time Bayesian estimation protocol estimates relaxation times within a few milliseconds, close to the decoherence timescale itself.
Our statistical analysis further suggests that some of these fast fluctuations arise from two-level systems switching at rates up to $\SI{10}{\hertz}$, four orders of magnitude faster than earlier reports.
These results redefine the timescales relevant for calibration in superconducting quantum processing units, establish a reference for rapid relaxation-rate characterization in device screening, and improve our understanding of fast relaxation dynamics. 

\end{abstract}
	\maketitle
	
	\section{Introduction} 
Superconducting qubits~\cite{Krantz2019, huang2020, Blais2021, gao2021practical, rasmussen2021} are among the main candidates for fault-tolerant quantum computation schemes, with quantum operations on these devices approaching error rates capable of demonstrating quantum error correction~\cite{Campbell2024, caune2024, acharya2025, Besedin2025}.
However, as the number of physical qubits increases in quantum processing units (QPUs), the QPU's performance is bounded by the lowest-performing outlier qubits~\cite{mohseni2024, Gao2025, wei2025low}. Identifying such outliers can be nontrivial, as time-dependent fluctuations in physical qubit parameters may alter which qubits qualify as outliers at any given moment. This is complicated by the physical mechanisms for these variations drifting over several different and competing timescales. In particular, the relaxation rate $\DecayRate$ of a superconducting qubit directly limits the fidelity of quantum operations~\cite{Malley2015}. $\DecayRate$ fluctuates unpredictably in the time domain~\cite{Siddiqi2021, Murray2021,mcrae2021, bejanin2021, Carroll2022, thorbeck2023, Zanuz2024, kim2024error,weeden2025} and also as a function of the qubit frequency~\cite{Siddiqi2021, Murray2021,mcrae2021, bejanin2021, Carroll2022, chen2025,weeden2025}. One of the major contributions to energy relaxation in state-of-the-art superconducting qubits is frequently attributed to their (semi-)resonant interaction with environmental two-level system (TLS) defects~\cite{muller2019, Siddiqi2021,Murray2021}, though their detailed microscopic origins remain unknown. It has been shown that TLS frequencies drift over repeated cooldowns, whereas the overall number of TLSs does not~\cite{Shalibo2010, Zanuz2024}. 

Previous works in transmon qubits have resolved $\DecayRate$ with a time resolution of seconds~\cite{Klimov2018,Schloer2019, Carroll2022} or minutes~\cite{Mueller2015, Burnett2019}. More efficient and scalable estimation methods are needed to (i) probe previously unexplored sub-second regimes of $\DecayRate$ dynamics, which is necessary for understanding the underlying physics, and (ii) identify outlier qubits and time-dependent fluctuations~\cite{muller2019, Siddiqi2021, Murray2021, mcrae2021} in large QPUs to ensure fast and reliable characterization and error mitigation. Modern field-programmable gate array (FPGA) advancements have facilitated \textit{online} (during experimental data collection) Hamiltonian learning~\cite{granade2012, Gebhart2023}, which is a useful tool to probe drifts in qubit parameters through real-time estimation~\cite{Gebhart2023, spiecker2023, Reuer2023, Arshad2024, Berritta2024a, dumoulin2024silicon, vora2024ml, Berritta2024b, Park2025, Berritta2025}.  

In this work, we investigate fast fluctuations of the relaxation rates of two long-lived transmon qubits (with $T_1 \equiv 1/\DecayRate\approx\SI{0.17}{\milli\second}$ measured over several hours) on millisecond timescales, almost comparable to the relaxation times themselves. We use a commercial controller with an integrated FPGA that leverages single-shot readout and performs real-time Bayesian estimation of the relaxation time, overcoming the sampling limitations of the traditional method mentioned below by two orders of magnitude. Our adaptive estimation allows the controller to investigate the stochastic behavior of the relaxation rate on unprecedentedly short timescales, revealing events where the relaxation time switches by almost an order of magnitude on the timescale of tens of milliseconds, instead of minutes or hours~\cite{Siddiqi2021, Murray2021, mcrae2021}.
We analyze the fluctuations of $\DecayRate$ using the power spectral density and Allan deviation~\cite{van1982} and find that a simple Lorentzian model describes the observed fluctuations, allowing us to extract TLS switching rates as fast as $\SI{10}{\hertz}$, $10^4\times$ faster than previously reported, and to monitor changes in the dominating TLS environment with sub-minute time resolution.

 To appreciate the advantages of our approach, it is useful to recall the traditional protocol. The most common method for estimating $\DecayRate$ consists of initializing the qubit to the excited state and measuring its state projectively after a fixed (\emph{nonadaptive}) waiting time. For each waiting time $\tau_{\text{wait}}$, the measurement outcome is averaged over many repetitions, and this is repeated for all waiting times. The fraction of times the system is measured in the excited state as a function of waiting time is then fitted to an exponential decay $\propto e^{-\DecayRate \tau_{\text{wait}}}$. The drawbacks of such a nonadaptive method are that (i) it is not optimally efficient in terms of experiment time, (ii) prior estimates of the qubit's $\DecayRate$ are not used to track its subsequent temporal fluctuations, and (iii) implementing curve fitting directly on an FPGA is challenging due to the available numerical precision, which prevents interleaving such estimations with qubit operations in real time.
 
 Bayesian parameter estimation~\cite{Gebhart2023, Kurchin2024} instead is a natural approach to implementing real-time optimization techniques~\cite{Gebhart2023, Arshad2024, Berritta2024a, Berritta2024b, Park2025, Berritta2025} compatible with low-latency control hardware. By using an onboard probability distribution of the parameter estimate, the optimal experimental settings for each subsequent probing cycle may be chosen \emph{adaptively} so as to maximize the knowledge obtained from every (single-shot) measurement. In this work the technique is validated with shot-by-shot interleaved estimations of the relaxation time using our Bayesian method and the traditional approach.  
 
 Our scheme sets a new reference for fast QPU characterization of relaxation rates, for probing relaxation dynamics at previously unexplored time resolution, and is well suited for stable interleaved QPU execution in the presence of $\DecayRate$ fluctuations.
	\section{Setup and protocol} 
\begin{figure}
\centering
\includegraphics{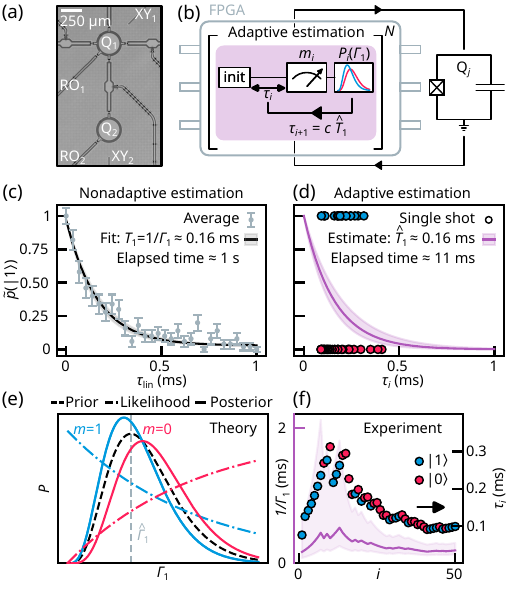}
\caption{\textbf{Device and Bayesian adaptive decay-rate estimation.} 
(a) Optical micrograph of transmon qubits ($\text{Q}_1$ and $\text{Q}_2$) nominally identical to the ones used in this work. Each transmon ($\text{Q}_j$) is individually controlled by microwave pulses (XY$_j$) and read out through independent resonators (RO$_j$).
(b) Experimental scheme for adaptively estimating the qubit decay rate $\DecayRate$ (purple box) on the FPGA in real time over $N$ probe cycles. In each cycle, labeled $i$, the controller initializes $\text{Q}_j$ to the excited state (init), waits a time $\tau_i$ adaptively chosen based on $\hat{T}_1 \equiv 1/ \langle \DecayRate\rangle$ from the previous Bayesian distribution ${\cal P}_{i-1}(\DecayRate)$, then updates the probability distribution ${\cal P}_i(\DecayRate)$ based on the measurement outcome $m_i$.
(c) Example of a nonadaptive estimation of the relaxation time. From the normalized fraction $ \Tilde{p}(\ket{1}) $ of excited states as a function of linearly stepped probing waiting times $\tau_{\text{lin}}$, $\DecayRate$ is estimated by an exponential fit. The total elapsed time is $\approx \SI{1}{\second}$.
(d) Example of an adaptive estimation of $\hat{T}_1$ by Bayesian statistics implemented on the controller. Each circle is a single-shot measurement outcome $\ket{0}$ (red) or $\ket{1}$ (blue) which updates the current estimate of $\hat{T}_1$ and the subsequent adaptive waiting time $\tau_i$. The total elapsed time for the entire estimation is $\approx \SI{11}{\milli\second}$.
(e) Evolution of the probability distribution ${\cal P}$ during each probe cycle $i$ of the estimation algorithm. The current estimate is $\hat{\mathit{\Gamma}}_1$ (gray dashed line). On each probing cycle, the estimate of $\hat{\mathit{\Gamma}}_1$ is updated according to two possible likelihood functions (dot–dashed lines), multiplied by the prior distribution (dashed). This yields the posterior distribution (solid), whose estimate is shifted left or right depending on the measurement outcome, while the uncertainty is reduced on average.
(f) Convergence of ${\cal P}$ as a function of the $i^{\text{th}}$ probe cycle. The line shows the resulting estimate $\hat{T}_1$ and the shaded area marks the 90\% credible interval.
As in panel (d), each circle is a single-shot outcome $m_i$ with corresponding waiting time $\tau_{i}$.
}
\label{fig:fig1}
\end{figure}
We use a 5-qubit superconducting array with tunable couplers operated at the mixing-chamber stage (below $\SI{10}{\milli\kelvin}$) of a dilution refrigerator. The device design and fabrication is similar to Ref.~\cite{Warren2023}. We implement the decay-rate estimation protocol on two of the five available qubits. In the main text, we focus on a transmon qubit ($\text{Q}_1$) of frequency $\approx \SI{4.13}{\giga\hertz}$ and anharmonicity $\approx\SI{-210}{\mega\hertz}$, as shown in Fig.~\ref{fig:fig1}(a). The results from another qubit ($\text{Q}_2$) in the same device are presented in the Supplemental Material~\cite{supplementary}. In our qubits, the expected Purcell-limited relaxation time is $\approx\SI{0.8}{\milli\second}$. 
A commercial controller (Quantum Machines OPX1000) applies microwave pulses for qubit control and single-shot readout, using dedicated XY control lines and readout resonators respectively [See Fig.~\ref{fig:fig1}(a)]~\cite{Krantz2019,huang2020, Blais2021, gao2021practical, rasmussen2021}. 

The fluctuating parameter $\DecayRate$ is estimated on the FPGA from the probing sequence shown in Fig.~\ref{fig:fig1}(b). For each probe cycle (labeled $1\leq i\leq N$), the qubit is initialized to the ground state $\ket{0}$ via active reset~\cite{riste2012} and is then brought to the excited state $\ket{1}$ using an X$_{\pi}$ pulse. After waiting for an adaptive waiting time $\tau_i$, the controller assigns the qubit state as ground ($m_i=0$) or excited ($m_i=1$) by thresholding the demodulated dispersive readout signal. The Bayesian probability distribution $\mathcal{P}_i(\DecayRate)$, detailed in Sec.~\ref{sec:bayes}, is updated based on $m_i$. The controller then uses the updated estimate of $\hat{T}_{1}\equiv 1/\langle \DecayRate\rangle$ based on $\mathcal{P}_i(\DecayRate)$ and selects a new waiting time $\tau_{i+1} = c\, \hat{T}_{1}$, with $c$ fixed, for the subsequent probing cycle.
 
To contrast our estimation method with the standard (nonadaptive) approach, we first present in Fig.~\ref{fig:fig1}(c) a ``standard $T_1$ experiment''. For clarity, the measurement outcomes of Fig.~\ref{fig:fig1}(c-d) ignore readout errors.
Specifically, in Fig.~\ref{fig:fig1}(c) we plot the normalized measured fractions of excited states, $ \tilde{p}(\ket{1}) $ as a function of the linear waiting time $\tau_{\rm lin}$, along with the corresponding exponential fit, yielding a decay constant of $ T_1 = 1/\DecayRate = \SI{165(15)}{\micro\second} $. This experiment is based on 1,890 single-shot measurements and spans approximately one second, comparable to the method used in Ref.~\cite{Klimov2018}. 

In contrast, in Figure~\ref{fig:fig1}(d), we present a representative run of our adaptive approach. The controller uses 50 single-shot measurement outcomes (experiment performed the day after the nonadaptive one), each of which is used to update the probability distribution on the controller iteratively. The updated distribution is then used to calculate an adaptive waiting time for the subsequent probing shot. After 50 single-shot measurements, the estimated value is $ \hat{T}_1 \approx \SI{159}{\micro\second} $, with a 68\% credible interval (CI) of $[131, 202]\,\SI{}{\micro\second}$, and a total elapsed time of only $ \SI{11}{\milli\second}$ $(\approx69\,\hat{T}_1)$. The purple curve illustrates the corresponding exponential decay, the shaded area indicating the CI.%

We see that the resulting uncertainty from the adaptive method is slightly larger than the nonadaptive case, while the estimator $\hat{T}_1$ was obtained with a total estimation time that is two orders of magnitude shorter. Moreover, in Fig.~\ref{fig:fig1}(c), the chosen values of waiting times within the interval $\left[1,1000\right]\,\SI{}{\micro\second}$ are appropriate for the estimated relaxation time in this specific example. However, if $T_1$ were to fluctuate drastically, a nonadaptive estimation on the same grid would result in a much greater uncertainty. The adaptive scheme is much more robust to these fluctuations as the waiting times are selected dynamically (for further details on how the adaptive method outperforms the nonadaptive one, see the Supplemental Material~\cite{supplementary}).

\section{Bayesian estimation} \label{sec:bayes}

We now describe our efficient and adaptive Bayesian estimation method for the decay rate implemented on the controller. The crux of the protocol is that all information about the current estimate of the probability distribution, $\mathcal{P}(\DecayRate)$, is stored at any time, dynamically and with only a few parameters, on the controller. This on-controller parametrization allows each Bayesian update of $\mathcal{P}(\DecayRate)$ to take only  $\approx\SI{2.2}{\micro\second}$ (cf.~the update time of $\approx \SI{50}{\micro\second}$ in Ref.~\cite{Arshad2024}, which used particle filtering for estimating decoherence rates in a nitrogen-vacancy center). Section~\ref{sec:results} presents our main results on $\DecayRate$ estimation, which remain accessible without reference to the following implementation details.

An overview of our Bayesian estimation method was described in the previous section, however we reiterate to be explicit as to what is executed at every step of the protocol and its approximations. The estimation method involves a probing cycle where the qubit is first initialized in the excited state $\ket{1}$. This initialization is followed by a waiting time $\tau$ after which the state of the qubit is measured using dispersive readout. Since the fluctuations of $\DecayRate(t)$ tend to be dominated by low-frequency noise~\cite{Siddiqi2021, Murray2021}, we approximate $\DecayRate(t)$ to be quasistatic on the scale of tens of probing cycles (i.e., a few ms). In the following, we thus drop the time dependence of $\DecayRate(t)$ for ease of notation. 

We assume that after the initialization the probability of measuring an outcome $m\in \{0,1\}$ corresponding to the states $|0\rangle$ and $|1\rangle$ is given by the likelihood function [dot-dashed lines in Fig.~\ref{fig:fig1}(e)]
\begin{equation}
P(m|\DecayRate, \tau) = 1-m - (-1)^m [\beta + (1-\alpha-\beta)e^{-\DecayRate \tau}],
\label{eq:likelihood}
\end{equation}
where $\DecayRate$ is the parameter we want the controller to estimate, and $\alpha$ and $\beta$ are misclassification probabilities for measuring $\ket{0}$ when at the beginning of the measurement the true state is $\ket{1}$ and measuring $\ket{1}$ when the true state is $\ket{0}$, respectively~\footnote{Our model assumes that an exponential decay well approximates a relaxation time experiment. However, this assumption breaks down in the presence of strong coupling between the qubit and a TLS, which induces oscillations in the excited-state return probability. If the oscillation frequency is known, the protocol can be easily adapted by sampling at the maxima (or minima) of these oscillations. Otherwise, the model must be extended for multi-parameter Bayesian estimation~\cite{valeri2023, Arshad2024}. Such strong coupling regimes have not been observed in Q$_1$ and Q$_2$ during our experiments.}. 

In the quasistatic approximation, by Bayes' rule, one obtains
\begin{equation} \label{eq:posterior}
{\cal P}_{i+1}(\DecayRate) \propto {\cal P}_i(\DecayRate)P(m_{i+1}|\DecayRate,\tau_{i+1}),
\end{equation}	
where the prior ${\cal P}_i(\DecayRate)$ describes the probability distribution for $\DecayRate$ after the $i^{\text{th}}$ probing cycle, which depends on all the previously used waiting times and measurement outcomes, and the posterior ${\cal P}_{i+1}(\DecayRate)$ describes the distribution after the subsequent cycle. 

 A simple parametrization of probability distributions is generally favorable in Bayesian approaches for on-FPGA optimization and computational speed, as demonstrated in Ref.~\cite{Berritta2025}.
 In typical experiments, the distribution of estimated $\DecayRate$'s fits well to a Gaussian~\cite{Burnett2019}, which suggests approximating the prior and posterior in Eq.~(\ref{eq:posterior}) by a Gaussian for all $i$~\cite{Ferrie2013,Berritta2025}.
 However, combining a Gaussian prior with the exponential likelihood function in Eq.~(\ref{eq:likelihood}) results in a posterior that is not well approximated by a Gaussian distribution, especially when the spread of the Gaussian becomes comparable to its mean, yielding also the unphysical ingredient of significant probability for negative values of $\DecayRate$.
 
Instead, we find that the gamma distribution is a convenient choice,
 	\begin{equation}
	\mathcal{P}_i(\DecayRate |k_i,\theta_i)=\frac{\theta_i^{k_i}}{\GammaFunction(k_i)}\DecayRate^{k_i-1}e^{-\theta_i\DecayRate},
	\label{eq:prior}
	\end{equation}
 where $k_i$ is the shape parameter, $\theta_i$ is the scale parameter, and $\GammaFunction(k)$ is the gamma function [for all positive integers $\GammaFunction(k)= (k-1)!$]. An example of a gamma-distributed prior with $k_i=5$ is shown in Fig.~\ref{fig:fig1}(e) (black dashed line).
 A gamma-distributed prior yields a posterior that remains exactly gamma-distributed in the absence of state-preparation and measurement errors $(\alpha = \beta = 0)$ when $m = 1$. Therefore, we expect that the posterior [see solid lines in Fig.\ref{fig:fig1}(e)] obtained via Eq.~(\ref{eq:posterior}) using the prior in Eq.~(\ref{eq:prior}) is in all relevant cases still approximately gamma-distributed. The gamma distribution is also a convenient choice for the case when the standard deviation becomes comparable to the mean, since by definition it has zero weight at negative $\DecayRate$. For standard deviations small compared to the mean ($k_i \gg 1$) the gamma distribution approaches a Gaussian distribution.
 
 In the following we will thus always approximate the probability distribution for $\DecayRate$ after the $i^{\text{th}}$ probing cycle with the gamma distribution $\mathcal{P}_i(\DecayRate |k_i,\theta_i)$. We emphasize that while the gamma distribution is often associated with stochastic waiting times, in this work it is not linked to the physical origin of $\DecayRate$~\footnote{A gamma distribution models the statistics of the sum of independent decay processes, where the random variable represents the total decay time. In that sense, it is naturally related to decay dynamics. However, in our case, we measure the state $m$ rather than the actual decay time $\tau$. As a result, the relevant statistics are binomial rather than the exponential distribution typically associated with lifetime studies. Therefore, the appearance of the gamma distribution in this work is not due to physical origins, but rather because it closely resembles the true posterior.}. Rather, it is implemented in the controller for its mathematical convenience and since it provides a good approximation of a Gaussian distribution after sufficiently many measurements. In summary, the use of gamma distributions is advantageous as it requires two parameters ($k_i,\,\theta_i$) only, and once multiplied by the exponential likelihood function (\ref{eq:posterior}), the posterior remains close to a gamma distribution. The controller thus only needs to keep track of two parameters after each measurement, which reduces the time the Bayesian update takes to $\approx\SI{2.2}{\micro\second}$, as mentioned above.
	
 Two steps are required to implement the estimation in the controller. The first step is to determine an adaptive waiting time $\tau$ based on the prior distribution. Then, one must approximate the resulting posterior ${\cal P}_{i+1}(\DecayRate)$ to a gamma distribution. This can be done using equations implemented directly on the controller in real time. 
\begin{figure*}
    \centering
    \includegraphics{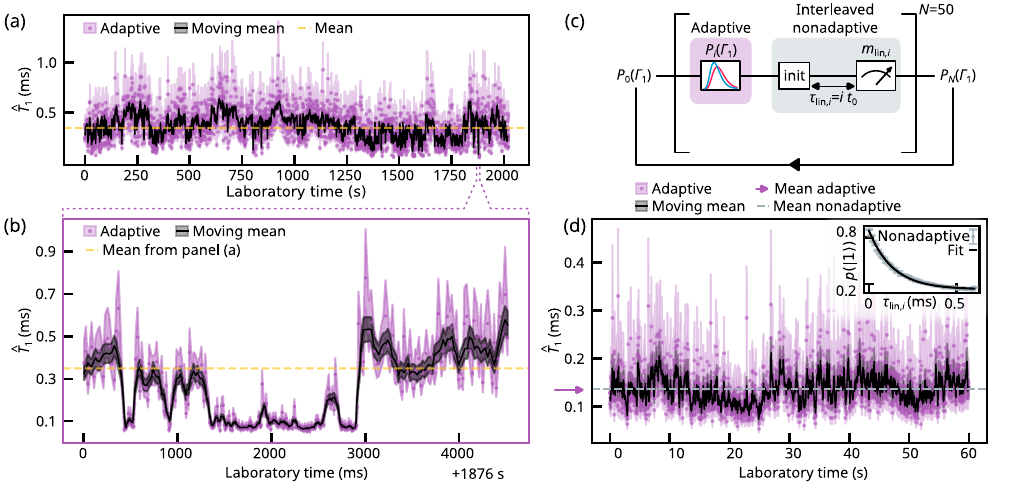}
    \caption{\textbf{Protocol for tracking and validation of the decay-rate fluctuations by adaptive estimation on the controller.} 
(a) Experimental results for the adaptive tracking protocol with $N= 100$ to estimate $\hat{T}_1$ (purple dots, downsampled by $D=30$) and its 68\% credible interval (shaded area). Each purple point in this plot required an average estimation time of $\approx\SI{20}{\milli\second}$. The dashed yellow line indicates the mean value of all estimates $\overline{T}_1\approx \SI{350}{\micro\second}$. The black line is a moving mean over 100 samples and the black shaded area is its 68\% confidence interval (see main text).
(b) Estimated $\hat{T}_1$ at $\approx \SI{1876}{\second}$ of panel~(a) (see purple dashed lines), where $\hat{T}_1$ shows telegraphic switching with timescales on the order of tens or hundreds of milliseconds.	
(c)
The interleaved estimation sequence for $\DecayRate$ used to validate the adaptive protocol.
Each of the $N$ probe cycles, labeled $i$, consists of parts contributing to the adaptive (purple) and nonadaptive (gray) estimates. Each adaptive probe cycle is followed by the nonadaptive part of the cycle, where the qubit is again initialized in the excited state, the wait time is fixed to $\tau_{\text{lin},i}=i\tau_0$, and the measurement outcome is stored for offline post-processing.
After the $N$ probe cycles, the final adaptively obtained distribution ${\cal P}_N(\DecayRate)$ is saved.
(d) Experimental results for the adaptive tracking protocol, interleaved with nonadaptive measurements. Main panel: The estimate $\hat{T}_1 $ (purple dots) and 68\% credible interval (shaded area) corresponding to the final probability distribution $\mathcal{P}_{50}(\DecayRate)$ of the 2,000 adaptive estimates performed during the $\approx \SI{60}{\second}$ of the experiment. The black line shows a moving mean over 5 samples and the black shaded area is its 68\% confidence interval. The purple arrow indicates the mean of all the adaptive estimates $\hat{T}_1$. The dashed line is the value extracted from the fit shown in the inset. Inset: Experimental results for the nonadaptive estimate using linearly sampled waiting time $\tau_{\text{lin}, i}$. Error bars represent the standard error.
}
\label{fig:fig3}
\end{figure*}
 \paragraph{Adaptive waiting time.} Working within a Bayesian framework, a common approach is to choose parameters that minimize the expected value of a quantity that measures the inaccuracy of the $\DecayRate$ estimate, such as the variance or Shannon information of its probability distribution, after a future measurement is obtained. This works well as long as one can find an analytical expression for the optimal experiment based on the chosen metric \footnote{Thus for adaptive methods, the Shannon information is rarely used in practice, as it tends to be computationally intractable.}. Here it turns out that even a simple metric like the expected posterior variance does not give a simple analytical update rule for choosing $\tau$. Therefore, in our Bayesian framework, we make a similar heuristic choice for the adaptive parameter $\tau$ as in Ref.~\cite{Arshad2024}:
 In each cycle, the controller uses the current estimate $\hat{T}_1\equiv 1/\langle \DecayRate\rangle = \theta/k$ based on the prior distribution and uses the adaptive waiting time:
 \begin{equation}
 \tau_{i+1} = c\, \hat{T}_{1,i}, 
 \end{equation}
 where $c$ is fixed in each experiment and depends on the qubit cycle idle time $t$ (e.g., initialization, readout), and measurement error rates $\alpha$ and $\beta$. The optimal choice for the coefficient $c$ is theoretically bound to the interval $c\in\left(0,1.59\right)$ and is chosen based on binomial statistics, to reduce the estimation time locally and uncertainty given the current knowledge of $\DecayRate$ (see the Supplemental Material~\cite{supplementary}). 
 
 \paragraph{Posterior approximation.} The prior $ {\cal P}_i(\DecayRate)$ in Eq.~(\ref{eq:posterior}) is assumed to be a gamma distribution as given by Eq.~(\ref{eq:prior}), illustrated by the black dashed line in Fig.~\ref{fig:fig1}(e). After measuring $m_{i+1}=\left\{0,1\right\}$, the posterior distribution 	${\cal P}_{i+1}(\DecayRate)$ is obtained by inserting Eqs.~(\ref{eq:likelihood},\,\ref{eq:prior}) in Eq.~(\ref{eq:posterior}). Since the posterior distribution is usually not an exact gamma distribution, we use the method of moments and approximate ${\cal P}_{i+1}(\DecayRate)$ with the gamma distribution which has the same mean $\mu_{i+1} = \text{E}[\DecayRate|m_{i+1}, \nu_{i+1}]$ and variance $\sigma^2_{i+1} = \text{E}[\DecayRate^2|m_{i+1}, \nu_{i+1}] - \text{E}[\DecayRate|m_{i+1}, \nu_{i+1}]^2$ computed over ${\cal P}_{i+1}(\DecayRate)$, where we use the notation $\nu_i \equiv (k_i,\theta_i,\tau_{i+1})$~\footnote{For $(\alpha = \beta = 0)$, we have $\text{E}[\DecayRate|m_i, \nu_i]= k / \theta$ and $\text{E}[\DecayRate^2|m_i, \nu_i]=(k_i+k_i^2)/\theta_i^2$ (see the Supplemental Material~\cite{supplementary} for the case $\alpha,\beta \neq 0$).}. 
 The required gamma distribution then has the parameters 
\begin{subequations}
 \begin{align}
 \theta_{i+1}^{-1} & = f_m(k_i+1,\theta_i,\tau_{i+1})-f_m(k_i,\theta_i,\tau_{i+1}),\\ 
 k_{i+1}^{-1} & =\frac{f_m(k_i+1,\theta_i,\tau_{i+1})}{f_m(k_i,\theta_i,\tau_{i+1})}-1, 
 \end{align}
\end{subequations}
where
\begin{subequations}
\begin{align}
 f_0(k,\theta,\tau) = {} & {} \frac{k}{\theta} \frac{1-\beta-(1-\alpha-\beta)\left(\frac{\theta}{\theta+\tau}\right)^{k+1} }{1-\beta-(1-\alpha-\beta)\left(\frac{\theta}{\theta+\tau}\right)^{k} },\\
 f_1(k,\theta,\tau) = {} & {} \frac{k}{\theta} \frac{\beta+(1-\alpha-\beta)\left(\frac{\theta}{\theta+\tau}\right)^{k+1} }{\beta+(1-\alpha-\beta)\left(\frac{\theta}{\theta+\tau}\right)^{k} }.
\end{align}
\end{subequations}
The approximated ${\cal P}_{i+1}(\DecayRate)$ becomes then the new prior and the controller repeats this scheme in total $N$ times per estimation repetition to obtain a sufficiently narrow distribution~\footnote{
We perform a numerical approximation of the Kullback–Leibler divergence ($D_{\mathrm{KL}}$) to estimate how much the posterior distribution differs from our gamma-distribution approximation. 
The largest deviation occurs for the worst-case outcome $m=1$ together with a small prior shape parameter, typically $k=3$. In this regime the posterior is strongly skewed, and the approximation error reaches a maximum of roughly $D_{\mathrm{KL}} \approx 0.09$--$0.10$ at intermediate ratios $\tau_{i+1}/\theta\gtrsim 1$. The accuracy improves systematically with increasing $k$. For example, at $k = 10$ the peak divergence drops below $0.006$, and for $k = 20$ it falls below $0.001$ across the entire range of $\tau_{i+1}/\theta$. See the Supplemental Material~\cite{supplementary} for further details on the validity of the approximation, also with respect to a normal distribution.}, during which $\DecayRate$ is assumed to be quasistatic. If one were to relax the quasistatic approximation, the optimum choice of $N$ would depend on details of the $\DecayRate$ drift and the estimation efficiency~\cite{benestad2024efficient, supplementary}.

In Fig.~\ref{fig:fig1}(f) we illustrate (i) the resulting evolution of ${\cal P}_i(\DecayRate)$ as a function of the measurement number $i$ for one representative estimation sequence, and (ii) the corresponding waiting times $\tau_i$ and measurement outcomes $m_i$. The misclassification probabilities $\alpha = 0.11$ and $\beta = 0.14$ are obtained by fitting an exponential decay from the nonadaptive method (e.g. see inset of Fig.~\ref{fig:fig3}(d) and Fig.~S3(c) of the Supplemental Material~\cite{supplementary}). The initial prior is defined by $(k_0, \,\theta_0) = (3,\,\SI{450}{\micro\second})$, with $N = 50$ and $\tau \approx 0.51/\langle \DecayRate\rangle$~\footnote{Due to the numerical precision of the controller, the actual ratio between $\tau_{i+1}$ and $\hat{T}_{1,i}$ varies by approximately 10\% in the vast majority of probe cycles.}, based on previously measured relaxation rate fluctuations. The purple line shows the estimates $\hat{T}_1$ and the shaded area indicates their 90\% credible interval, narrowing $\hat{T}_1$ down from $\SI{168}{\micro\second}\; (90\% \text{ CI: } [80, 630] \, \SI{}{\micro\second})$ to $\SI{191}{\micro\second}\; (90\% \text{ CI: } [133, 301] \, \SI{}{\micro\second})$ after 50 single-shot measurements. The user can pre-define the number of single-shot measurements based on the required target uncertainty, which is traded off against estimation speed. 
Upon measuring $m_i = 1$ (blue circle), the estimate $\hat{T}_{1,i}$ increases compared to its previous value, and the subsequent waiting time $\tau_{i+1}$ also increases. In contrast, after measuring $m_i = 0$ (red circle), $\hat{T}_{1,i}$ decreases, while $\tau_{i+1}$ decreases on average.

\section{Results} \label{sec:results}

\subsection{Tracking high-frequency fluctuations by adaptive estimation}

We now apply our fast adaptive method to characterize the qubit at a previously unexplored temporal resolution. We first task the controller to acquire a time trace of $\hat{T}_1$ by the adaptive estimation, with $N=100$ probe cycles~\footnote{Each probe cycle includes a $\SI{2.5}{\micro\second}$ readout period, followed by an approximately $\SI{8}{\micro\second}$ wait time to allow the resonator to cool down and fully deplete any residual photons from the readout pulse. Each Bayesian update takes $\approx\SI{2.2}{\micro\second}$. The controller is programmed to start with an initial gamma distribution prior with $(k_0, \,\theta_0)=(3,\,\SI{450}{\micro\second})$. With this in mind, the adaptive waiting time in each cycle is set to $\tau_{i+1} \approx 0.51 \,\hat{T}_{1,i}$ (see the Supplemental Material~\cite{supplementary} for a motivation of the choice $c=0.51$), where $\hat{T}_{1,i}$ is the estimator for $T_1$ after cycle $i$.} for each estimation repetition.  We plot the estimated $\hat{T}_1$ on the controller in Fig.~\ref{fig:fig3}(a). The 68\% confidence interval of the moving mean is computed using the standard error of the standard deviations of the 5 Bayesian posterior distributions: The standard error is estimated by first calculating the moving mean of the posterior standard deviations and then dividing by the square root of the window size. The confidence bounds are then determined as the moving mean of the estimated $\hat{T}_1$ values $\pm 1$ standard error under normality assumption.
 
In Fig.~\ref{fig:fig3}(b), we plot the estimated $\hat{T}_1$ around $\SI{1878}{\second}$ of panel~(a) (other windows are presented in the Supplemental Material~\cite{supplementary}). As shown, $\hat{T}_1$ switches between $\gtrapprox\SI{500}{\micro\second}$ and $\approx\SI{100}{\micro\second}$ on a timescale of tens to hundreds of milliseconds.  We emphasize that the purple points are \emph{independent} of each other since at the beginning of each estimation repetition the prior distribution is reset to $\mathcal{P}_0(\DecayRate)$.

The fluctuations in panels Fig.~\ref{fig:fig3}(a-b) exhibit telegraphic noise with multiple stable points, which could be explained by the qubit interacting with an ensemble of TLSs~\cite{muller2019, Siddiqi2021, Murray2021} changing due to spectral diffusion~\cite{Klimov2018,weeden2025} or background ionizing radiation~\cite{thorbeck2023}. We highlight that such fast fluctuations would not be measurable using the standard nonadaptive method, which has previously reported sampling times of a few seconds~\cite{Klimov2018,Schloer2019, Carroll2022}. The dwell times, on the order of tens of milliseconds, are consistent with the state-switching dynamics observed in a TLS strongly coupled to a superconducting qubit used as a detector~\cite{Meissner2018}. Our sampling interval, which is two orders of magnitude faster than previous works~\cite{Klimov2018,Schloer2019, Carroll2022}, reveals that significant $T_1$ fluctuations can occur at least one order of magnitude faster than previously reported in transmon qubits~\cite{Mueller2015, Klimov2018, Burnett2019, Schloer2019, Carroll2022}. 

The estimation time is only tens of times longer than the average $T_1$ and can be integrated into adaptive quantum control strategies to error-mitigate the performance of QPUs. One may imagine interleaving the estimation protocol with quantum circuit operations, executing them only when the estimated qubit relaxation time $T_1$ exceeds a user-defined threshold to maintain a target gate fidelity. Since $T_1$ fluctuates over time, instead of running quantum circuits continuously, the system can pause its execution when the relaxation rate is too high, thus improving the overall QPU fidelity. Since both the interleaved estimation and qubit operations are performed on the same qubit, there exists a nontrivial relationship between the correlation time of the fluctuations being tracked, the efficiency of the estimation process, and the duration required for coherent operations between estimations. 
Overall, the results presented in this section demonstrate that our Bayesian estimation protocol performs real-time tracking of the decay rate of a superconducting qubit. 

\subsection{Protocol validation by interleaved adaptive and nonadaptive estimations} 
Next, we validate the protocol by programming the controller to perform interleaved measurements with the nonadaptive method to verify that it correctly identifies the decay rate $\DecayRate$. The fluctuating parameter $\DecayRate$ is estimated from the probing sequence shown in Fig.~\ref{fig:fig3}(c), where each probe cycle $i$ begins with an adaptive cycle [see Fig.~\ref{fig:fig1}(b)] and is interleaved cycle-by-cycle with the nonadaptive one (gray)~\footnote{Here $(k_0, \,\theta_0)=(3,\, \SI{450}{\micro\second})$, $\alpha = 0.11$, $\beta = 0.14$. Considering also that the nonadaptive $\tau_{\text{lin},i}$ is linearly stepped up to $\approx \SI{600}{\micro\second}$, the average idle time from the point of view of the adaptive scheme is $t\approx\SI{345}{\micro\second}$, mostly covered by the average $\langle \tau_{\text{lin},i}\rangle \approx \SI{300}{\micro\second}$. The adaptive waiting time in each cycle is then set to $\tau_{i+1} \approx 0.98 \,\hat{T}_{1,i}$ (see the Supplemental Material~\cite{supplementary}).}, the outcomes $m_{\text{lin},i}$ of which are stored.
 
In Fig.~\ref{fig:fig3}(d) we characterize the experimentally found final posterior probability distribution ${\cal P}_{50}(\DecayRate)$ from 2,000 subsequent adaptive estimations. The 68\% confidence interval of the moving mean is computed as in Fig.~\ref{fig:fig3}(a). In post-processing we calculate the 50 averages $\langle m_{\text{lin},i} \rangle$ and fit them to an exponentially decaying curve, see the inset panel of Fig.~\ref{fig:fig3}(d), which yields $T_1 = \SI{136.7(2.2)}{\micro\second}$ (dashed gray line in the main panel). The 2,000 adaptive estimations give on average $\overline{T}_1 = \SI{135.0(0.9)}{\micro\second}$ (purple arrow in the main panel), computed from the mean and standard error of the adaptive time trace shown in the main panel of Fig.~\ref{fig:fig3}(d). The main result is that the two values agree, and the reduced uncertainty of the adaptive method results from the narrowing of the prior distribution and the adaptive waiting time chosen for the experiment. 
 
The controller performs the same interleaved estimation procedure on another qubit for over 3 minutes ($\text{Q}_2$, located on the same chip, see the Supplemental Material~\cite{supplementary}). In that case, the nonadaptive method yields $ T_1 = \SI{178(1.7)}{\micro\second} $, which is again in good agreement with the mean adaptive estimate $ \overline{T}_1 = \SI{182.63(0.55)}{\micro\second} $. Compared to $\text{Q}_1$, we attribute the slightly larger discrepancy between these values to residual $T_1$ fluctuations occurring between the interleaved probing cycles.
 
\subsection{Power spectral density and Allan deviation} \label{sec:allan_psd}
\begin{figure*}
		\centering
	\includegraphics{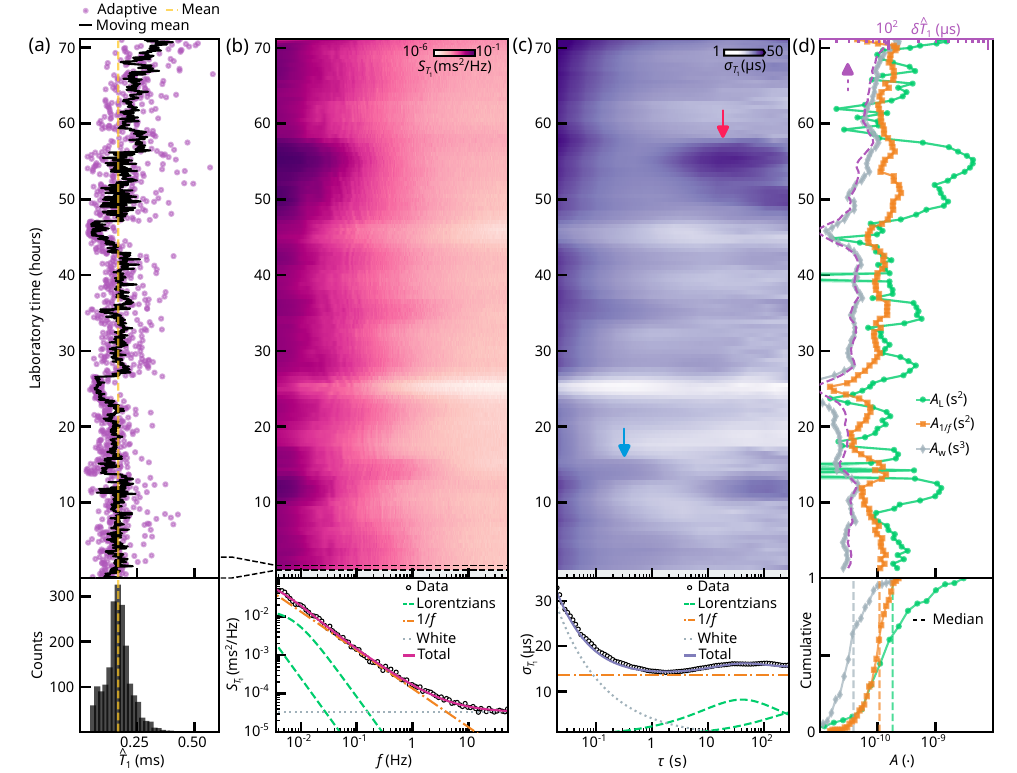}
	\caption{\textbf{Frequency and time domain analysis of $\hat{T}_1$ fluctuations on a 72-hour timescale} 
(a) Estimated $\hat{T}_1$ (purple dots, downsampled by $D=30,000$) as a function of laboratory time by real-time adaptive tracking with $N= 49$ and sampling speed of $\approx\SI{7}{\milli\second}$ over 72 hours. The dashed yellow line shows the mean value of all estimates $\overline{T}_1\approx \SI{168}{\micro\second}$. The black line is a moving mean with a window of size 20,000 over the original estimates. Lower panel: Histogram of the moving mean.
(b-c) Frequency and time domain analysis of the $\hat{T}_1$ fluctuations shown in panel~(a), obtained from 2.8-hour running windows with 80\% overlap (black dashed lines): 
(b) Power spectral density. Lower panel: PSD of the full time trace.
(c) Allan deviation on a logarithmic scale; arrows denote the zoom-in regions of Fig.~\ref{fig:fig5}. Lower panel: Allan deviation of the full time trace.
(d) Amplitudes of the Lorentzian ($A_{\text{L}}$), 1/$f$ ($A_{1/f}$), and white noise ($A_\text{w}$) contributions, extracted from the full time trace, by a simultaneous PSD and Allan deviation fit to the analytical formulas from Table~\ref{tab:noise}. The purple dashed line is the standard deviation $\delta \hat{T}_1$ (see main text) of the Bayesian posterior distribution. Lower panel: Cumulative histogram of the fitted $A_i$ and their medians (dashed lines).
}
		\label{fig:fig4}
\end{figure*}

To gain further insight into the physics of the fast fluctuations we observe, we calculate the power spectral density (PSD) and the Allan deviation~\cite{van1982, Burnett2019, ye2024} of a time trace of $\hat{T}_1$  acquired over 72 hours. The entire trace consists of $\approx3.82\times 10^7$ samples in total and is shown in Fig.~\ref{fig:fig4}(a). The controller uses $N=49$ single-shot measurements per estimation repetition, with settings similar to those in Fig.~\ref{fig:fig3}(b)~\footnote{$\left(k_0,\,\theta_0\right) = \left(3,\SI{600}{\micro\second}\right)$, $\alpha=\beta = 0.12$, and $c \approx 0.53$.}.

\begin{figure}
 \centering
 \includegraphics{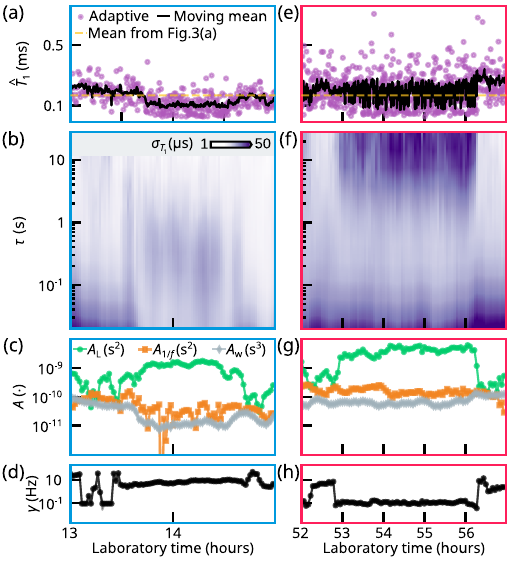}\caption{\textbf{Time domain analysis of $\hat{T}_1$ fluctuations on sub-minute observation times.} Zoom-ins of the regions indicated by the arrows in Fig.~\ref{fig:fig4}(c).
 (a) Estimated $\hat{T}_1$ (purple dots, downsampled by $D=3,000$) as a function of laboratory time. The dashed yellow line represents the mean value of all estimates $\overline{T}_1$ of Fig.~\ref{fig:fig4}(a). The black line is a moving mean with a window size of 3,000 over the original estimates. 
 (b) Allan deviation in logarithmic scale of the $\hat{T}_1$ time trace with a running window of about 7 minutes and 80\% overlap. 
 (c) Amplitudes of the Lorentzian ($A_\text{L}$), $1/f$ ($A_{1/f}$), and white ($A_\text{w}$) noise contributions in each interval extracted from panel~(a) by a simultaneous fit to the Allan deviation and PSD using analytical models from Table~\ref{tab:noise}. 
 (d) Fitted switching rate $\gamma\approx \SI{10}{\hertz}$ over more than one hour of the Lorentzian component.
 (e-h) Same as (a-d) with a running window of 17 minutes and 80\% overlap. The switching rate fit in panel~(h) reveals a stable Lorentzian process with $\gamma\approx \SI{100}{\milli\hertz}$ for more than 3 hours. 
 }
 \label{fig:fig5}
\end{figure}

We compute the PSD [Fig.~\ref{fig:fig4}(b)] and Allan deviation [Fig.~\ref{fig:fig4}(c)] with a running 2.8-hour-long window (with 80\% overlap) in the top panels and for the full trace in the lower panels. Compared to the controller’s fast sampling period of $\approx\SI{7}{\milli\second}$, the relatively long window enables us to resolve Lorentzian processes in the Allan deviation within hundreds of seconds of observation time. Later, in Fig.~\ref{fig:fig5}, we use narrower windows to highlight faster dynamics.

In Fig.~\ref{fig:fig4}(b), top panel, we show the power spectral density $S_{\hat{T}_1}(f)$. The PSD of $\hat{T}_1(t)$ is defined as the Fourier transform of its autocorrelation function: 
\begin{equation}
S_{\hat{T}_1}(f) = \int_{-\infty}^{+\infty} \langle \hat{T}_1(t) \hat{T}_1(t + \tau)\rangle e^{-2\pi i  f \tau} \text{d}\tau. 
\end{equation}

The PSD exhibits increased amplitude at low frequencies over some laboratory times, but attributing this behavior to a TLS is not straightforward, as even a few TLSs can produce a smooth, featureless $1/f$ spectrum that conceals individual contributions. 
In contrast, the Allan deviation quantifies how much a signal switches on average over an observation time $\tau$. If an individual TLS induces a random telegraph signal with a characteristic switching time, the Allan deviation of this signal exhibits a peak close to the average switching time of the TLS~\cite{Burnett2019, ye2024}. 

For our qubit parameter $ \hat{T}_1(t)$, the Allan deviation is defined as:
\begin{equation}
\sigma_{\hat{T}_1}(\tau) = \sqrt{\frac{1}{2} \left\langle \left( \overline{T}_1(t+\tau, \tau) - \overline{T}_1(t, \tau) \right)^2 \right\rangle}, 
\end{equation}
where $\overline{T}_1(t, \tau) = \frac{1}{\tau} \int_{t}^{t+\tau} \hat{T}_1(t') \text{d}t'$ is the average over an interval of duration $ \tau $ and $\langle \ldots \rangle$ is the average over $t$. In Fig.~\ref{fig:fig4}(c), top panel, we plot $\sigma_{\hat{T}_1}(\tau)$ which is high at short observation times ($\tau\lessapprox \SI{100}{\milli\second}$) due to high-frequency noise and sampling uncertainty. As $\tau$ increases, $\sigma_{\hat{T}_1}(\tau)$ decreases as the noise averages out. At longer $\tau$, switching in $\hat{T}_1$ shows up as an increase in $\sigma_{\hat{T}_1}(\tau)$, which is clearly observed around $\tau$ in the range of tens of seconds at laboratory times of 11, 50, and 55 (red arrow) hours. Another striking feature is the peak at 14 hours (blue arrow) in the sub-second timescale. All the peaks shown in Fig.~\ref{fig:fig4}(c) are a clear signature of a Lorentzian noise process, as no power-law noise source can reproduce them, and they correspond to the rises in the PSD at smaller frequencies. The peaks of the Allan deviations are indications of individual TLSs that move in and out of resonance with the qubit~\cite{bejanin2021, Carroll2022}. To our knowledge, these features have not been previously observed at such short observation timescales and are accessible here because of the fast and adaptive estimation protocol. The PSD and Allan deviation are lower at 25 and 45 hours of laboratory time, as a shorter $T_1$ generally corresponds to smaller fluctuations~\cite{Mueller2015,Klimov2018,Schloer2019, Burnett2019, bal2024systematic, kono2024}. We tentatively attribute the shorter $\hat{T}_{1} \approx \SI{0.1}{\milli\second}$ to one or several TLSs being close to resonance with the qubit.

\begin{table}
 \centering
 \caption{Power spectral density (PSD) and Allan deviation models for different noise processes. The amplitudes $A_i$ and switching rate $\gamma$ are free parameters for the simultaneous fit of the PSD and Allan deviation.}
 \begin{tabular}{l c c}
 \toprule
 \toprule
 \multicolumn{1}{l}{\textbf{Noise}} & \multicolumn{1}{c}{PSD ${S_{\hat{T}_1}(f)}$} & \multicolumn{1}{c}{Allan deviation $\sigma_{\hat{T}_1}(\tau)$} \\
 \midrule
 White & $A_\text{w}$ & $\sqrt{A_\text{w}/\tau}$ \\
 \midrule
 $1/f$ & $A_{1/f}f^{-1}$ & $\sqrt{2A_{1/f}\ln 2}$ \\
 \midrule
 Lorentzian & $\frac{4A_{\text{L}} \gamma}{\gamma^2 + (2\pi f)^2}$ & $\sqrt{A_{\text{L}}} \gamma\tau
 \sqrt{2\gamma \tau + 1 -(e^{-\gamma \tau} - 2)^2 }$ \\ 
 \bottomrule
 \bottomrule
 \end{tabular}
 \label{tab:noise}
\end{table}

To quantify the impact of Lorentzian-type noise on the measured $\hat{T}_1$ fluctuations, it is common practice to fit the full time trace to a model including white noise, $1/f$ noise, and one or more Lorentzian noise sources~\cite{Burnett2019}. We note that even if two Lorentzian processes fit the full trace, different TLS dynamics occur over laboratory time and it is actually sufficient for the fit to use only one Lorentzian in a shorter window slice. In the following, we present both analyses to support our claim.

In the bottom panels of Fig.~\ref{fig:fig4}(b-c) we plot the PSD and Allan deviation from the full trace shown in panel~(a). Both curves are simultaneously fitted with a sum of white, $1/f$, and Lorentzian noise components, using the analytical expressions from Table~\ref{tab:noise}~\cite{van1982, Burnett2019}. Each noise process has a corresponding amplitude coefficient, $A_i$, which is a free parameter for the Allan deviation $\sigma_{\hat{T}_1}(\tau)$ and the power spectral density $S_{\hat{T}_1}(f)$ simultaneously. The Lorentzian fit also includes the switching rate $\gamma$ as a free parameter. To fit the full trace we use two Lorentzian processes and the fit parameters are given by $A_\text{w} = \SI{3.30(0.06)e-5}{\second\cubed}$, $A_{1/f} = \SI{1.36(0.02)e-4}{\second\squared}$, $A_{\text{L,1}} = \SI{1.8(0.1)e-4}{\second\squared}$, $\gamma_1 = \SI{46(5)}{\milli\hertz}$, $A_{\text{L,2}} = \SI{1.0(0.3)e-4}{\second\squared}$, and $\gamma_2 = \SI{2(1)}{\milli\hertz}$. 

Now, instead of the full trace, we fit each 2.8-hour-long time window of the top panels of Fig.~\ref{fig:fig4}(b-c) to show that only one Lorentzian process is sufficient to fit the data instead of two by selecting a shorter window. We plot the amplitudes of each noise process contribution in Fig.~\ref{fig:fig4}(d) and their cumulative histogram in the bottom panel. In the main panel of (d), the 68\% confidence intervals of the fit parameters are smaller than the plotted symbols and are calculated as the square roots of the diagonal elements of the covariance matrix associated with the fit. The characteristic switching rate of the fitted Lorentzian component is shown in the Supplemental Material~\cite{supplementary}, including the close agreement between the model and the data. The $1/f$ noise likely originates from the qubit interacting with an ensemble of TLSs~\cite{muller2019, Siddiqi2021, Murray2021} which dominate the low-frequency fluctuations of $T_1$ as observed from our fit. The presence of distinct Lorentzian components supports the existence of a few strongly coupled fluctuators, consistent with the telegraphic switching observed in Fig.~\ref{fig:fig3}(b). 

In Fig.~\ref{fig:fig4}(d) we also plot the standard deviation $\delta \hat{T}_1 \approx \langle \DecayRate \rangle ^{-2} \delta \DecayRate = \theta k^{-3/2}$ (purple dashed line) from the posterior distribution computed on the controller. We note that $\delta \hat{T}_1$ correlates very well with $A_\text{w}$ (gray diamonds), which is the fitted amplitude of the white noise contribution according to our model. The strong correlation between the white noise power and the posterior standard deviation suggests that $\delta \hat{T}_1$ indeed reflects the true uncertainty in the estimate. In other words, if the white noise scales with $\delta \hat{T}_1$ above a known noise floor, it is strong indirect evidence that the Bayesian posterior's $\delta \hat{T}_1$ reliably captures the actual estimation error. To our knowledge, such correlation in real-time estimation methods has not been reported before.

To quantitatively extract the occurrence rate of large telegraphic switches in $\hat{T}_1$ on tens of milliseconds timescales [Fig.~\ref{fig:fig3}(b)], we split the 72-hour-long time trace of Fig.~\ref{fig:fig4}(a) into independent 200-ms-long intervals. We find that $\approx 2.6\%$ of intervals exhibit on average changes in $T_1$ greater than $\SI{100}{\micro\second}$, roughly one event every $\SI{7.7}{\second}$ (see the Supplemental Material~\cite{supplementary}).

As mentioned above, from Fig.~\ref{fig:fig4}(c) we observe dominant Lorentzian noise contributions for instance in the regions marked by the arrows. To highlight the fast noise dynamics pointed at by the blue arrow, we zoom in on the 2-hour time period around hour 14 and plot the $\hat{T}_1$ estimates in Fig.~\ref{fig:fig5}(a). To increase the temporal resolution, we analyze 7-minute intervals with 80\% overlap and compute the corresponding Allan deviation for each, as shown in Fig.~\ref{fig:fig5}(b) with peaks around laboratory time of 14 hours and within sub-second observation time $\tau$. The fitted parameters are presented in Fig.~\ref{fig:fig5}(c) alongside the switching rate $\gamma$ in Fig.~\ref{fig:fig5}(d).

The fitted amplitudes reveal distinct periods during which Lorentzian fluctuators dominate the $T_1$ dynamics. In particular, the analysis identifies a more than one-hour-long window dominated by a single fluctuator with a relatively stable switching rate of $\gamma \approx \SI{10}{\hertz}$, consistent with the timescales observed by Ref.~\cite{Meissner2018}. Such fast dynamics would be extremely challenging to probe with traditional nonadaptive methods. 

We also focus on the region marked by the red arrow of Fig.~\ref{fig:fig4}(c) over 5 hours, see the corresponding $\hat{T}_1$ estimates in Fig.~\ref{fig:fig5}(e). Its Allan deviation is shown in Fig.~\ref{fig:fig5}(f), along with the fit parameters in panels (g) and (h). In this time span, the dominating TLS has a stable switching rate of about~$\SI{100}{\milli\hertz}$, slower than the $\SI{10}{\hertz}$ presented in Fig.~\ref{fig:fig5}(d) but still two orders of magnitude faster than what is reported, for instance, in Ref.~\cite{Burnett2019}. In panels (c,\,d,\,g,\,h), the 68\% confidence intervals of the fit parameters are smaller than the plotted symbols and are calculated as in Fig.~\ref{fig:fig4}(d).

In this section we have demonstrated how our adaptive Bayesian estimation method, about two orders of magnitude faster than conventional nonadaptive methods~\cite{Klimov2018, Schloer2019, Carroll2022}, confirms dominant Lorentzian noise processes in transmon qubits in a previously unexplored regime of quickly fluctuating relaxation times.
	
	\section{Conclusions and outlook} 
 This work presents the experimental demonstration of an adaptive Bayesian estimation protocol for the decay rate of two fixed-frequency transmon qubits. The scheme probes sub-millisecond relaxation times using only tens of single-shot measurements, with total acquisition times of a few milliseconds. Our approach achieves rapid adaptive estimation of the decay rate by integrating real-time Bayesian estimation with FPGA-based feedback control. The protocol has been validated by interleaved measurements with the standard nonadaptive method of extracting the decay rate from fitting an exponential decay curve. 
 
 The main results of our work are (i) our fast protocol improves the state-of-the-art estimation speed by two orders of magnitude without compromising accuracy~\cite{Klimov2018,Schloer2019, Carroll2022}, (ii) it reveals clear telegraphic changes in $\DecayRate$ by almost one order of magnitude [Fig.~\ref{fig:fig3}(b)], with dwell times of tens of milliseconds, instead of minutes or hours, most likely caused by interaction with an environmental bath of TLSs, and (iii) temporal and spectral analyses of the estimated fluctuations are consistent with dominant Lorentzian processes, from which switching rates as fast as $\SI{100}{\milli\hertz}$ and $\SI{10}{\hertz}$ are resolved in Fig.~\ref{fig:fig5}, four orders of magnitude faster than previously reported. The estimation scheme is a powerful probe of dominant TLSs that switch in and out of resonance with the qubit on timescales of tens of seconds. Such timescales would be extremely difficult to observe with standard nonadaptive methods with the precision reported here, and they redefine the timescales relevant for calibration in superconducting QPUs, which traditionally operate on minute-to-hour cycles.
 
 Our protocol uncovers new perspectives on materials characterization. High-throughput qubit screening targeting fast fluctuations previously required hours to accumulate sufficient statistics~\cite{Burnett2019, mcrae2021}. While long acquisition times remain necessary for characterizing slow drifts, our scheme collects fast fluctuations statistics within seconds. Average relaxation times of $T_1 \gtrapprox \SI{100}{\micro\second}$~\cite{tuokkola2025, dane2025} have been shown in superconducting qubits, however achieving such results uniformly across a large wafer~\cite{mohseni2024} and over time remains challenging. The overall performance is limited by the worst $T_1$ values, and $T_1$ tails can be improved through better fabrication that mitigates drops in $T_1$. Our protocol provides a useful tool for qubit benchmarking and process control, enabling rapid optimization of superconducting qubits fabrication. Since the protocol employs only single-qubit gates, it can be readily extended to multiple qubits simultaneously, and its adaptiveness makes it appealing for the characterization of large qubit arrays with unknown decay rates.

 In future work, the protocol could be interleaved with dephasing rate measurements~\cite{Schloer2019} using existing real-time estimation techniques~\cite{Arshad2024, Berritta2025}. Potential correlations between decay and dephasing rates may provide insights into the microscopic origins of intrinsic decoherence channels at higher frequencies in Josephson junction qubits. Additionally, the estimation could be combined with spectral manipulation of TLSs via applied electric fields~\cite{lisenfeld2019, bilmes2020, kim2024error, chen2025, dane2025} or strain~\cite{grabovskij2012, lisenfeld2019}. Our adaptive Bayesian technique could be employed for simultaneous relaxation time estimations across multiple qubits~\cite{Carroll2022, kono2024}, to investigate temporal and spatial correlations at significantly higher frequency bandwidths than previously achieved. Spectral dynamics could also be explored in fixed-frequency transmons by implementing off-resonant microwave tones to drive AC Stark shifts~\cite{Carroll2022, chen2025}. Additionally, our method could be applied for the rapid detection of qubit decay induced by gamma and cosmic rays~\cite{Martinis2021, Siddiqi2021, Murray2021, harrington2024}. 
 
 Potential modifications to the protocol include relaxing the assumption of single-shot readout~\cite{Arshad2024}, or mitigating state preparation and measurement errors by repeating probe cycles with the same waiting times~\cite{Sergeevich2011, Cappellaro2012}, at the cost of a slower estimation rate. Furthermore, the estimation could be optimized by terminating it once a target total measurement time or desired uncertainty is reached, rather than using a fixed number of single-shot measurements. 

 Fluctuations in the decay rate degrade the stability of QPUs, introduce uncertainty in coherence benchmarking, and hinder process optimization for superconducting qubits. Quantum error correction demonstrations are predominantly limited by the ``worst'' outlier qubits in a given processor~\cite{mohseni2024, wei2025low}. Since state-of-the-art gate fidelities are limited by $\DecayRate$, the optimal gate duration depends on temporal variations of $\DecayRate$. The observed stochastic decay rate fluctuations suggest the following error mitigation approach: Continuous identification of the lowest-performing qubits, suggesting a shift from offline periodic recalibration every few hours to real-time adaptive recalibration at millisecond timescales for maintaining higher-fidelity gate operations in QPUs. More generally, the ability to monitor a qubit's relaxation rate in real time opens new opportunities for the dynamic optimization and routing of quantum algorithms. A recently estimated $\DecayRate$ value could be passed to a compiler's routing algorithm to determine the optimal mapping between virtual and physical qubits, thereby boosting the benchmarking or diagnostics of useful QPUs, and could be further extended to other platforms.
 
 Our adaptive Bayesian technique also offers an efficient and online Hamiltonian learning protocol for real-time estimation of decay rates beyond superconducting qubits. Our results support that TLSs are major contributors to rapid decay-rate fluctuations and a deeper understanding of TLSs is thus vital for further improving the performance of useful QPUs.

\section{Acknowledgments}
F.B., M.A.M., S.K., C.W.W., J.H. and M.K. gratefully acknowledge support from the Novo Nordisk Foundation (grant number NNF22SA0081175, the NNF Quantum Computing Programme), the Villum Foundation through a Villum Young Investigator grant (grant 37467), the European Union through an ERC Starting Grant (NovADePro, 101077479), the Innovation Fund Denmark (DanQ, Grant No. 2081-00013B), and the U.S. Army Research Office (Grant No. W911NF-22-1-0042).
F.B., J.~Benestad, J.A.K., J.D. and F.K. acknowledge funding from the European Union's Horizon 2020 research and innovation programme under grant agreements 101017733 (QuantERA II), 101204890 (HORIZON-MSCA-2024-PF-01) and EUREKA Eurostars 3 (ECHIDNA), and from the Dutch National Growth Fund (NGF) as part of the Quantum Delta NL programme and the Research Council of Norway (RCN) under INTFELLES-Project No.~333990.
O.K. received funding via the Innovation Fund Denmark for the project DIREC (9142-00001B).
The device was fabricated at Myfab Chalmers. The Chalmers team was funded by the Knut and Alice Wallenberg Foundation through the Wallenberg Center for Quantum Technology (WACQT) and the EU Flagship on Quantum Technology HORIZON-CL4-2022-QUANTUM-01-SGA project 101113946 OpenSuperQPlus100.
Any opinions, findings, conclusions, or recommendations expressed in this material are those of the author(s) and do not necessarily reflect the views of the Army Research Office, the US Government, the European Union, or the European Research Council. Neither the European Union nor the granting authority can be held responsible for them.

 \section{Author contributions}
F.B. led the measurements and data analysis, and wrote the manuscript with input from all authors. F.B., J.~Benestad, J.H., F.K., M.K. performed the experiment with theoretical contributions from J.A.K., O.K., J.D. M.A.M. and S.K. developed experimental infrastructure. C.W.W. designed the device, which was fabricated by E.H., A.N., I.A., A.O., J.~Bizn\'{a}rov\'{a}, M.R. under the supervision of A.F.R., J.~Bylander and G.T. The project was supervised by J.H., F.K. and M.K.
		
\section{Data availability}
The data that support the findings of this article are openly available~\cite{zenodo}.

	\bibliography{my_bibliography}
 
\end{document}


\beginsupplement
	
	
	
    \title{Supplemental Material for \\ ``Real-time adaptive tracking of fluctuating relaxation rates\\ in superconducting qubits''}
	
\author{Fabrizio~Berritta}
\altaffiliation{Current address: Research Laboratory of Electronics, Massachusetts Institute of Technology, Cambridge, MA 02139, USA} 
\email{fabrizio.berritta@mit.edu}
\affiliation{\QDevaffil}
\affiliation{\NNFaffil}	
 \author{Jacob~Benestad}
 \affiliation{\NTNUaffil}
 \author{Oswin~Krause}
	\affiliation{Department of Computer Science, University of Copenhagen, 2100 Copenhagen, Denmark}
 \author{Jan~A.~Krzywda}
 \affiliation{Lorentz Institute for Theoretical Physics \& Leiden Institute of Advanced Computer Science, Universiteit Leiden, 2311 EZ Leiden, The Netherlands}
 \author{Malthe~A.~Marciniak}
 \affiliation{\QDevaffil}	
 \affiliation{\NNFaffil}		
 \author{Svend~Kr{\o}jer}
 \affiliation{\QDevaffil}	
 \affiliation{\NNFaffil}	
 \author{Christopher~W.~Warren}
 \affiliation{\Chalmersaffil}
 \affiliation{\QDevaffil}	
 \affiliation{\NNFaffil}		
 \author{Emil~Hogedal}
 \affiliation{\Chalmersaffil}
 \author{Andreas~Nylander}
 \affiliation{\Chalmersaffil} 
 \author{Irshad~Ahmad}
 \affiliation{\Chalmersaffil} 
 \author{Amr~Osman}
 \affiliation{\Chalmersaffil}\author{Janka~Bizn\'{a}rov\'{a}}
 \affiliation{\Chalmersaffil}
 \author{Marcus~Rommel}
 \affiliation{\Chalmersaffil}
 \author{Anita~Fadavi~Roudsari}
 \affiliation{\Chalmersaffil}
 \author{Jonas~Bylander}
 \affiliation{\Chalmersaffil}
 \author{Giovanna~Tancredi}
 \affiliation{\Chalmersaffil} 
 \author{Jeroen~Danon}
	\affiliation{\NTNUaffil}
 \author{Jacob~Hastrup}
 \affiliation{\QDevaffil}	
 \affiliation{\NNFaffil}	
 \author{Ferdinand~Kuemmeth}
 \affiliation{\QDevaffil}	
 \affiliation{Institute of Experimental and Applied Physics, University of Regensburg, 93040 Regensburg, Germany}
 \affiliation{QDevil, Quantum Machines, 2750 Ballerup, Denmark}	
 \author{Morten~Kjaergaard}
 \email{mkjaergaard@nbi.ku.dk}
 \affiliation{\QDevaffil}	
 \affiliation{\NNFaffil}
	
	\date{February 13, 2026}
	\maketitle
	\tableofcontents
	
	\section{Experimental setup}
	\begin{figure*}[h]
		\centering
		\includegraphics[height=0.7\textheight,keepaspectratio] {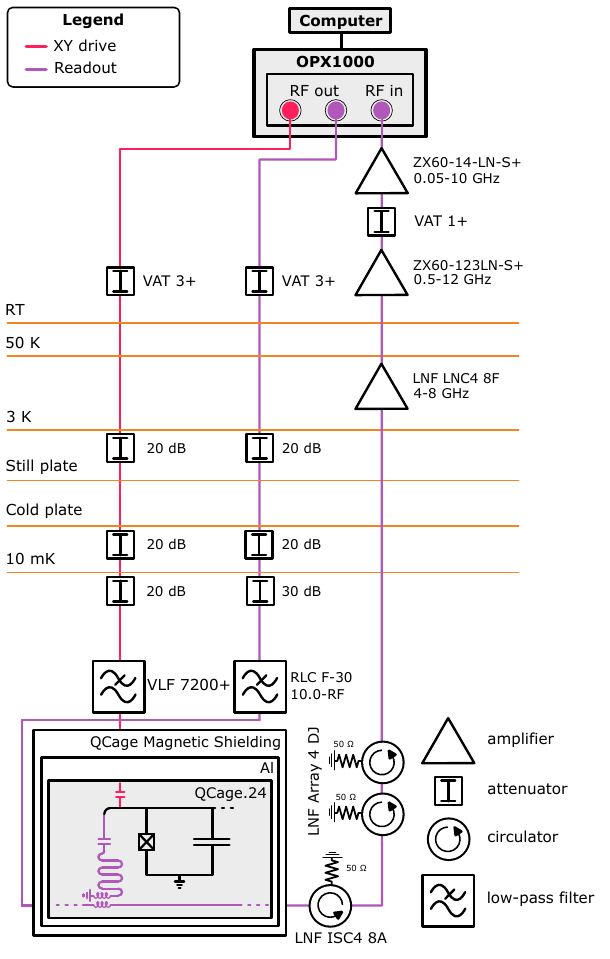}
		\caption[Experimental setup BF2]{\textbf{Experimental setup.} The cryostat is a Bluefors XLD400 dilution refrigerator with a base temperature lower than 10 mK. A Quantum Machines OPX1000 is used for the XY drive pulses and readout.}
		\label{fig:FigS1}
	\end{figure*}
The measurements are performed in a Bluefors XLD400 dilution refrigerator with a base temperature below $\SI{10}{\milli\kelvin}$ and the setup is sketched in Fig.~\ref{fig:FigS1}. The Quantum Machines OPX1000 is used for the XY control of the qubit and readout signal, and both microwave pulses are generated by direct digital synthesis. Each drive pulse is 60-ns long, with a Gaussian envelope of 12-ns standard deviation. The readout pulse is 2.5-$\SI{}{\micro\second}$ long, with a cosine rise and fall envelope of $\SI{20}{\nano\second}$ each. 

The OPX1000 includes real-time classical processing with fast analog feedback programmed in QUA software. The readout resonator linewidth $\kappa_\text{r}/(2\pi)\approx \SI{110}{\kilo\hertz}$, and the microwave readout tone, approximately $\SI{6.94}{\giga\hertz}$, is filtered and attenuated at room temperature by passive components. The readout tone is attenuated in the cryostat to remove excess thermal photons from higher-temperature stages, and low-pass filtered at the mixing chamber. The device sample is wirebonded to the printed circuit board (PCB) of a QCage.24 sample holder and the chip is suspended by four corners inside a cavity and clamped down by the PCB. The sample holder is placed inside a light-tight superconducting aluminum enclosure to reject stray microwave and infrared photons and it is then shielded magnetically with a QCage Magnetic Shielding. Both the sample holder and the magnetic shield are supplied by Quantum Machines. The transmitted signal from the feedline goes to a high-electron-mobility transistor amplifier thermally anchored at the  $\SI{3}{\kelvin}$ stage to amplify the readout signal further. At room temperature, the readout line is again amplified.

\clearpage

\section{Adaptive estimation in $\text{Q}_1$: additional examples of fast fluctuations}
\begin{figure*}
\centering
\includegraphics{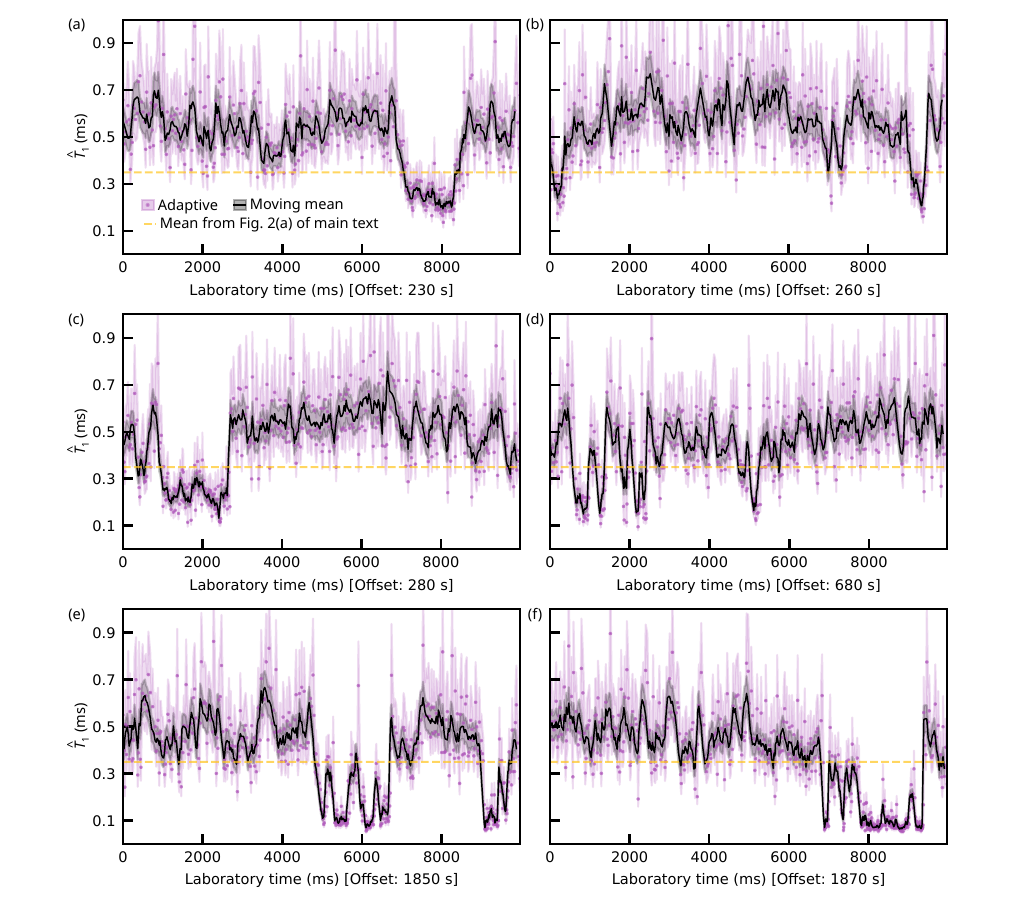}
\caption{\textbf{Tracking of the decay-rate fluctuations in Q$_1$ by adaptive estimation on the controller.}
        (a-f) Plotting not downsampled windows of Fig.~2(a) of the main text, where $\hat{T}_1$ shows fast switching with timescales on the order of tens or hundreds of milliseconds. Panel~(f) includes Fig.~2(b) of the main text.}
		\label{fig:FigS11}
	\end{figure*}
In Fig.~\ref{fig:FigS11} we plot different 10-second-long windows of the time trace shown in Fig.~2(a) of the main text. The fluctuations show telegraphic switching between high values of $\hat{T}_1 > \SI{500}{\micro\second}$ and low values $\hat{T}_1 \approx \SI{100}{\micro\second}$. 

\clearpage
\section{Decay-rate estimation in $\text{Q}_2$}	
\begin{figure*}
\centering
\includegraphics{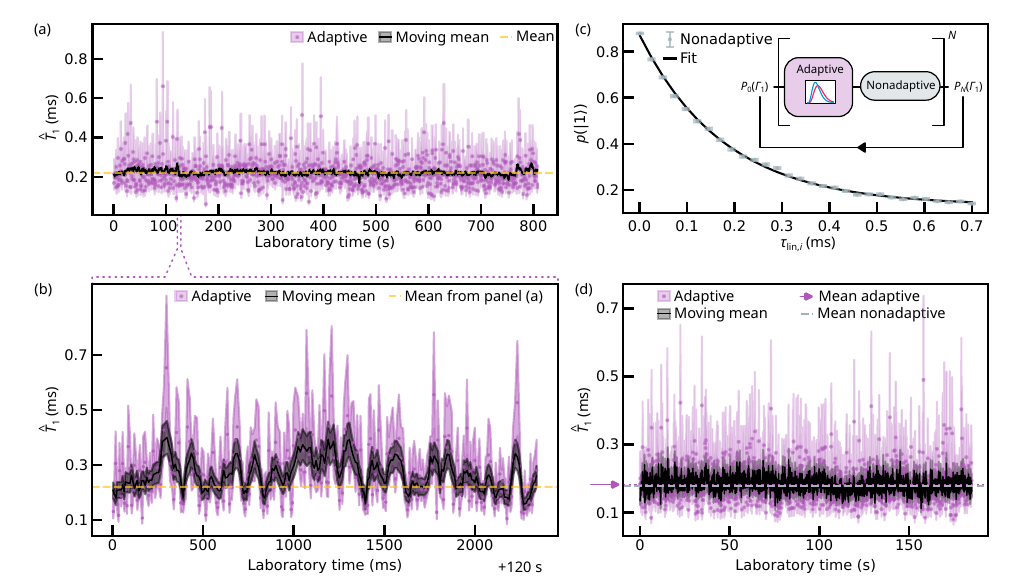}
\caption{\textbf{Tracking of the decay-rate fluctuations in Q$_2$ by adaptive estimation on the controller.} Reproducing Fig.~2 from the main text in Q$_2$.		 
        (a) Adaptive tracking with $N= 49$. Each purple point in this plot required an average measurement time of $\approx\SI{8}{\milli\second}$. The moving average is over 300 samples.
        (b) Zoom-in at $\SI{120}{\second}$ of panel~(a) (purple dashed lines), where $\hat T_1$ shows fast switching with timescales on the order of tens of ms.
        (c) Inset: Interleaved adaptive and nonadaptive estimation sequence. Main panel: Experimental results for the nonadaptive protocol.
        (d) Experimental results for the adaptive tracking protocol, interleaved with nonadaptive estimates in the inset panel. Main panel: $\hat T_1$ (purple dots) of the final probability distribution $P_{29}(\DecayRate)$ of the 10,000 estimates performed during the $\approx \SI{164}{\second}$ of the experiment. The black line is a moving average over 10 samples.
}
		\label{fig:FigS2}
	\end{figure*}
The aim of this section is to replicate the estimation protocols of Fig.~2 of the main text ($\text{Q}_1$) in another qubit ($\text{Q}_2$) in the same device for reproducibility. The qubit frequency of Q$_2$ is $\approx\SI{3.8}{\giga\hertz}$, the anharmonicity is $\approx\SI{-220}{\mega\hertz}$ and the associated readout frequency is $\SI{6.85}{\giga\hertz}$. The drive line configuration in the cryostat is nominally identical to what shown in Fig.~\ref{fig:FigS1} for $\text{Q}_1$.

\subsection{Adaptive estimation}
Here we perform the adaptive estimation similarly to Fig. 2(a) of the main text. For plotting clarity in Fig.~\ref{fig:FigS2}(a), we downsample by $D=100$ the original trace of 100,000 estimates over $\SI{8.06e2}{\second}$ of the experiment which used $N = 49$ single-shot measurements per estimation repetition. The black line represents a moving average of 300 samples. The dashed yellow line indicates the average $\overline{T}_1 \approx \SI{220}{\micro\second}$ over this dataset. 

\begin{figure*}
\centering
\includegraphics{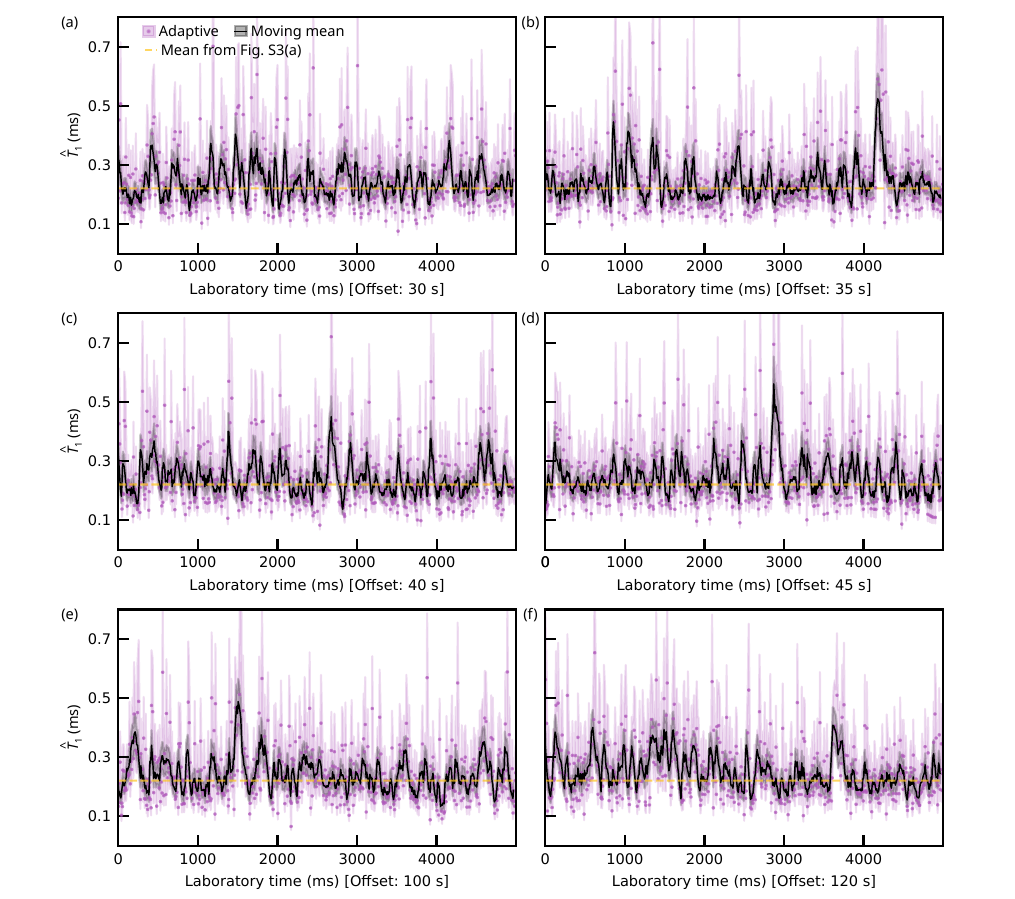}
\caption{\textbf{Adaptive estimation in $\text{Q}_2$: additional examples of fast fluctuations}
        (a-f) Plotting not downsampled windows of Fig.~\ref{fig:FigS2}(a), where $\hat{T}_1$ shows fast switching with timescales on the order of tens of milliseconds. Panel (f) includes Fig.~\ref{fig:FigS2}(b).}
		\label{fig:FigS10}
	\end{figure*}

In Fig.~\ref{fig:FigS2}(b), we replot the $\hat{T}_1$ timetrace of (a) around $\SI{120}{\second}$ (purple dashed lines) where $\hat{T}_1$ switches between $\gtrsim\SI{300}{\micro\second}$ and $\approx\SI{150}{\micro\second}$ on a timescale of tens of milliseconds. The moving average window consists of 5 samples. In Fig.~\ref{fig:FigS10} we plot different 10-second-long windows of the time trace shown in Fig.~\ref{fig:FigS2}(a). The fluctuations show telegraphic switching between high values of $\hat{T}_1 > \SI{400}{\micro\second}$ and low values $\hat{T}_1 \approx \SI{150}{\micro\second}$, with dwell times shorter than what measured in Q$_1$.

\subsection{Interleaved adaptive and nonadaptive estimations} 
    The controller interleaves the adaptive probe cycles with nonadaptive ones [inset Fig.~\ref{fig:FigS2}(c)], whose measurement outcomes $m_{\text{lin}, i}$ are averaged and fitted as shown in the main panel of Fig.~\ref{fig:FigS2}(c). The fit yields $T_1 = \SI{178.3(1.7)}{\micro\second}$.
    
    Regarding the duration of the probe cycle, the time used to read out the qubit is $\approx\SI{3}{\micro\second}$, and the time used to subsequently cool down the resonator again (to deplete it from any residual photons after the readout pulse) is $\approx \SI{8}{\micro\second}$. In this experiment $\alpha = 0.12, \beta = 0.13$ and the optimal waiting time in the $i$-th qubit cycles is given by $\tau_{i+1} \approx 1.0\, \hat T_{1,i}$.
    
    The controller is programmed to start with an initial gamma distribution prior with $(k_0, \theta_0)=(3,\, \SI{550}{\micro\second})$. 
    In Fig.~\ref{fig:FigS2}(d) we plot $\hat{T}_1$ and the black line is a moving average over 10 samples. 
    The estimation cycle is repeated $N=29$ times, while $\tau_{\text{lin}, i}$ is increased linearly from $\SI{1}{\micro\second}$ to $\SI{700}{\micro\second}$. The adaptive estimation gives $\hat{T}_1 = \SI{182.63(0.55)}{\micro\second}$ [purple arrow in Fig.~\ref{fig:FigS2}(d)], computed from the mean and standard error of the adaptive time trace. The value is close to the one extracted from the fit [$\SI{178.3(1.7)}{\micro\second}$] [gray dashed line in Fig.~\ref{fig:FigS2}(d)], with a slightly greater relative error compared to Fig.~2(d) of the main text. We attribute such small discrepancy to residual $T_1$ fluctuations between the interleaved probing cycles.

\clearpage
\section{Interleaved estimations with and without active reset}

\begin{figure*}
\centering
\includegraphics{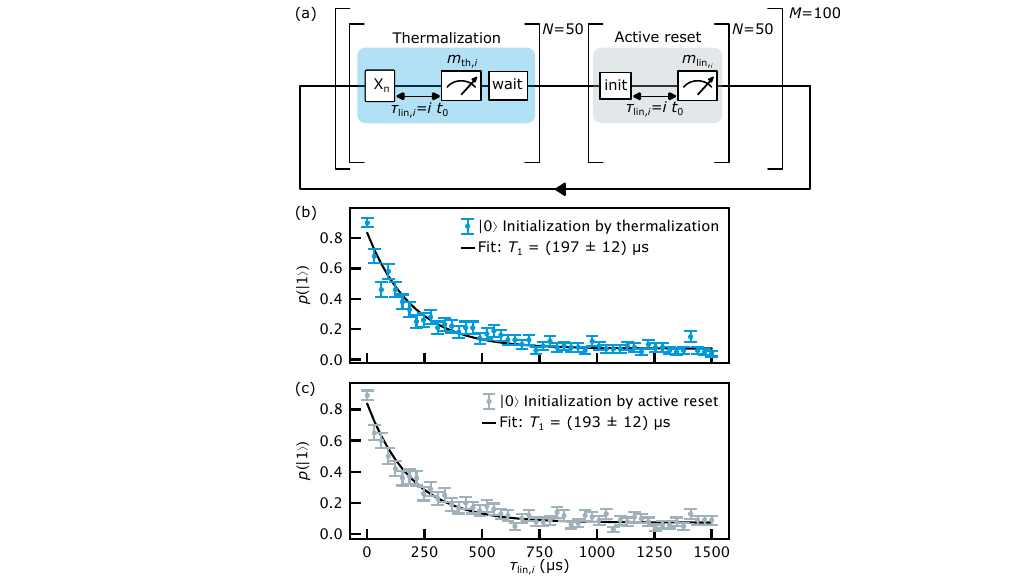}
\caption{\textbf{Interleaved $T_1$ measurement with and without active reset.}
(a) One loop (solid arrow) represents a single repetition of the protocol. In this experiment, the controller performs $ M = 100 $ repetitions. At the beginning of each repetition, for each of the $ N $ probe cycles (Thermalization), labeled $ i $, the controller initializes the qubit to the ground state by waiting $ \SI{1.5}{\milli\second} $ (wait), followed by an $ X_{\pi} $ pulse. It then waits for a linearly sampled time $ \tau_{\text{lin}, i} = i\tau_0 $ before measuring the qubit state ($ m_{\text{th}, i} $). To test whether active reset affects the estimated $ T_1 $ value, the first $ N = 50 $ probing cycles are followed by another $ N = 50 $ cycles (active reset), where the qubit is initialized to the excited state via active reset, followed by an $ X_{\pi} $ pulse. The waiting times $ \tau_{\text{lin}, i} = i\tau_0 $ are used before measurement ($ m_{\text{lin}, i} $).  
(b) Fraction of measured excited states $ m_{\text{th}, i} = 1 $ when the qubit is initialized to the ground state by thermalization.  
(c) Fraction of measured excited states $ m_{\text{lin}, i} = 1 $ when the qubit is initialized to the ground state by active reset and an $ X_{\pi} $ pulse. In panels (b) and (c) error bars represent the standard error from the experimental data.
}
		\label{fig:FigS3}
	\end{figure*}

Here we want to test whether active reset affects the $\DecayRate$ estimation by, for instance, quasiparticle pumping~\cite{gustavsson2016suppressing}. As shown in Fig.~\ref{fig:FigS3}(a), the experiment consists of $ M = 100 $ repetitions, where the controller performs $ N = 50 $ probe cycles with the qubit initialized to the ground state through thermalization by waiting $\SI{1.5}{\milli\second}$, followed by another $ N = 50 $ cycles where the qubit is actively reset to the excited state using a conditional $ X_{\pi} $ pulse. The measured fraction of excited states in each case is shown in panels Fig.\ref{fig:FigS3}(b) and (c), respectively. Panel~(b) presents the results for thermalization-based initialization which yields $T_1 = \SI{197(12)}{\micro\second}$, while panel~(c) yields $T_1 = \SI{193(12)}{\micro\second}$  when active reset is applied. As the two values agree within uncertainty, we conclude that active reset is unlikely to introduce systematic effects in the fluctuations of $T_1$.
\clearpage
\section{Optimal waiting time with binomial statistics}
\begin{figure*}
\centering
\includegraphics{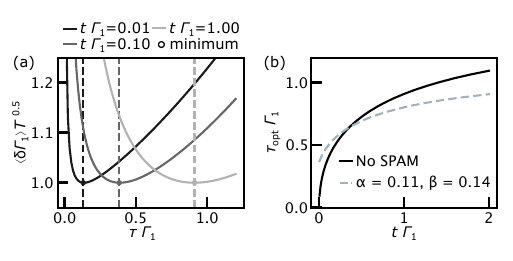}
\caption{\textbf{Optimal waiting time with binomial statistics}
(a) Expected normalized standard deviation $\langle \delta \hat{\mathit{\Gamma}}_1 \rangle \sqrt{T}$ when $\hat{\mathit{\Gamma}}_1$ is estimated based on a series of probing cycles with fixed $\tau$, the number of cycles being determined by a total time budget $T$. $\langle \delta \hat{\mathit{\Gamma}}_1 \rangle$ is plotted as a function of $\tau$ normalized by the relaxation rate $\DecayRate$. The curves differ by the idle time $t$ that takes into account the finite reset and measurement time, normalized by $\DecayRate$. The global minimum of each curve (dot) provides the optimal $\tau$ that minimizes the uncertainty.
(b) Optimal waiting time $\tau_{\text{opt}}$ as a function of the idle time $t$, both normalized by $\DecayRate$. The case without state and preparation measurement errors (SPAM) is the black curve where $\tau_{\text{opt}}$ approaches zero for sufficiently short idle times $t$. In the presence of SPAM (e.g. $\alpha = 0.11$ and $\beta = 0.14$, dashed-gray curve) $\tau_{\text{opt}}$ does not approach zero (see text). }
\label{fig:FigS4}
\end{figure*}
To estimate the relaxation rate as fast and precisely as possible, we first need to answer the following question: What is the locally optimal waiting time for estimating $\DecayRate$? By ``locally optimal" we mean within a greedy approach that minimizes both the experiment time and estimation uncertainty. To answer this question, we set aside the Bayesian framework and focus instead on the binomial statistics of the measurement outcomes for a fixed waiting time (i.e., the statistics of the number of successes for a number of independent trials), expanding on the main findings of Ref.~\cite{Leveraro2024}.

In any estimation protocol, there are generally two regimes of interest. The first regime is where the waiting time is negligible compared to the ``idle" time of each probing cycle (initialization, measurement, etc.). In this case, reducing the total measurement time involves optimizing the number of measurements. In the second regime, the waiting time is comparable to or exceeds the idle time. Here, one must determine whether it is more efficient to take a single measurement with an adaptively chosen long waiting time or to perform multiple shorter measurements within the same total measurement duration. In this section, we address this question, focusing on the latter regime where the waiting time is comparable to or much longer than the idle time, as in our experiment. The case of long idle times is discussed later.  

Our goal is to determine the waiting time $\tau$ that reduces the variance of the estimate of $\DecayRate$ most rapidly as a function of the total time $T$ the estimation takes. For this purpose, we consider $N$ consecutive single-shot measurements with waiting time $\tau$, the outcomes of which are expected to follow a binomial distribution $B(N,p)$ with $p=\beta+(1-\alpha-\beta)e^{-\DecayRate \tau}$ the probability to measure $\ket{1}$. Here $\DecayRate$ is the parameter we want the controller to estimate, and $\alpha$ and $\beta$ are misclassification probabilities for measuring $\ket{0}$ when at the beginning of the measurement the true state is $\ket{1}$ and measuring $\ket{1}$ when the true state is $\ket{0}$, respectively. 
When estimating $p$ from a set of measurement outcomes, the expected standard deviation for the estimator $\hat{p}$ due to the binomial measurement statistics is
\begin{equation}
    \langle \delta\hat{p}\rangle  = \sqrt{\frac{p-p^2}{N}}.
\end{equation}
It can be converted into a standard deviation for $\hat\DecayRate$ based on these $N$ measurement outcomes,
\begin{equation}
\langle \delta \hat{\mathit{\Gamma}}_1 \rangle  = \frac{\sqrt{\tau+t}}{(1-\alpha - \beta)\tau e^{-\DecayRate\tau} \sqrt{T}} \sqrt{\beta + (1-\alpha-\beta)e^{-\DecayRate\tau}}\sqrt{1-\beta - (1-\alpha-\beta)e^{-\DecayRate\tau}},
\label{eq:uncertainty_gamma_finite_time}
\end{equation}
where we used that $N = T/(\tau + t)$, with $t$ the required idle time delay between one measurement and the next, and $T$ the total time of the experiment.
We see that $\langle \delta \hat{\mathit{\Gamma}}_1 \rangle$ decays $\propto 1/\sqrt T$, and by
introducing the dimensionless variables $\tilde{\tau}=\DecayRate\tau$, $\tilde{t}=\DecayRate t$ and $\tilde{T}=\DecayRate T$ we derive an optimal measurement time $\tilde{\tau}^\ast = \tau_{\text{opt}}\DecayRate$ by solving
\begin{equation}
\label{opt_dimensionless}
    \frac{\partial}{\partial \tilde{\tau}}\left(\frac{\sqrt{\tilde{\tau}+\tilde{t}}}{\tilde{\tau} e^{-\tilde{\tau}} \sqrt{\tilde{T}}} \sqrt{[\beta + (1-\alpha-\beta)e^{-\tilde{\tau}}][1-\beta - (1-\alpha-\beta)e^{-\tilde{\tau}}]}\right)\equiv0.
\end{equation}
Except for some limiting cases, the solution to this equation must be found numerically. This gives us the specific value of the prefactor $c$ presented in the main text, which depends on SPAM coefficients $(\alpha,\beta)$ and qubit idle time $\tilde{t}$ normalized by $\DecayRate$. Note that the derived $\tau_{\text{opt}}$ is a function of the true $\DecayRate$, which is of course unknown. 

In Fig.~\ref{fig:FigS4}(a) we plot the uncertainty $\langle \delta \hat{\mathit{\Gamma}}_1 \rangle \sqrt{T}$, with minimum normalized to 1, for different values of idling time $t$ with $\alpha = \beta = 0$. Each curve shows a different global minimum corresponding to the optimal waiting time $\tau_{\text{opt}}$. In Fig.~\ref{fig:FigS4}(b) we plot the numerically found $\tau_{\text{opt}}$ as a function of the idle time $t$, both normalized by the decay rate $\DecayRate$, in the absence and presence of state preparation and measurement (SPAM) errors. We see that in absence of SPAM (black curve) as $t\rightarrow0$, the optimal waiting time $\tau_{\text{opt}} \rightarrow0$. This reflects the fact that as the idle time is reduced, it is better to perform more measurements with very short waiting times. In the presence of SPAM (dashed-gray curve), as in the experiment of Fig.~2 of the main text with $\alpha = 0.11$ and $\beta = 0.14$,  $\tau_{\text{opt}}$ does not approach zero. We notice this non-zero limit occurs only when $\alpha\neq 0$. 
On the other hand, by setting $\beta = 0$ and $t\rightarrow0$, we find one of the few analytical solutions of \eqref{opt_dimensionless}, which is:
\begin{equation}
    \tau_{\text{opt}}\DecayRate = 1 + \text{LambertW}\left(\frac{\alpha-1}{e}\right),
\end{equation}
where the Lambert W function, also known as the product logarithm, is defined as the inverse relation of $f(w)=we^{w}$.
\subsection{Adaptive waiting time implementation on the controller}
Since solving Eq.(\ref{opt_dimensionless}) numerically on the controller is not feasible given the complexity of the calculation and the available numerical accuracy on the FPGA, we program as fixed prefactor $c$ for the waiting time $\tau = c\, \hat{T}_1$ the one corresponding to $T_1=1/ \DecayRate = \SI{100}{\micro\second} $, on the order of magnitude of the observed $T_1$ in our device. In the experiment in Fig. 2(a)[(d)] of the main text, the average idle time $t\approx\SI{23.2}{\micro\second}\, [\SI{345}{\micro\second}]$. This choice deviates by a factor of 4.3 [1.5] in Fig.~2(a)[(d)] of the main text from the optimal waiting time of Fig.~\ref{fig:FigS4} considering the average value of the estimated $\overline{T}_1 \approx\SI{350}{\micro\second}\,[\SI{135}{\micro\second}]$. For the prefactor $c$, the choice of $T_1 = \SI{100}{\micro\second}$ remains a conservative choice to better estimate shorter $T_1$ values, as, in a QPU, the primary concern is whether the fidelity exceeds a user-defined threshold required for quantum error correction. Even though we do not observe a difference in performance for different values of $c$, we choose to better estimate worst-case parameter fluctuations rather than best-case values.
Due to the numerical precision of the controller, the actual ratio between $\tau_{i+1}$ and $\hat{T}_{1,i}$ varies slightly between probe cycles. A potential future improvement would be to precompute a lookup table of the numerically solved equation offline and upload it to the FPGA for real-time update of the prefactor $c$. 

\subsection{Optimal waiting time with the number of measurements}
Here we show how the standard deviation of our decay rate measurements scales with the \emph{number} of measurements performed during an experiment, neglecting that each measurement is time-dependent (in other words, we work in the limit $t\rightarrow + \infty$). The scheme is based on repeated measurements at the same sampling time $t$. These repeated measurements are averaged to produce a single data point, which is then used to estimate the parameter $\DecayRate$ of the exponential distribution:
\begin{equation}
    y(\tau) = \beta + (1-\alpha-\beta)\exp(-\DecayRate \tau)
\end{equation}
To estimate $\DecayRate$, given an estimate of the excited state return probability $y$ at some time $\tau$, we determine which $\tau$ gives the smallest variance on our estimate of $\DecayRate$. Here, $\tau$ corresponds to the probing time at which the qubit is measured. Let the point be $(y_0, \tau_0)$. The estimate of $\DecayRate$ is found by inverting $y(\tau)$:
\begin{equation}
    \DecayRate = -\frac{1}{\tau_0}\log\left(\frac{y_0-\beta}{1-\alpha-\beta}\right).
    \label{eq:decay_rate}
\end{equation}
The uncertainty in $\DecayRate$ is determined by differentiating Eq.~(\ref{eq:decay_rate}):
\begin{equation}
    \delta\DecayRate = \left| \frac{\mathrm{d}}{\mathrm{d}y_0} \left( -\frac{1}{\tau_0}\log\left[\frac{y_0-\beta}{1-\alpha-\beta}\right] \right) \delta y_0 \right|.
\end{equation}
Calculating the derivative, we find:
\begin{equation}
    \delta\DecayRate = \frac{1}{(y_0-\beta) \tau_0} \delta y_0.
\end{equation}
Assuming we are measuring a qubit where $N$ is the number of single-shot measurements, the uncertainty in $y$ at $(\tau_0, y_0)$ due to binomial statistics is:
\begin{equation}
    \langle \delta \hat{y}_0 \rangle = \sqrt{\frac{y_0 - y_0^2}{N}}
\end{equation}
We observe that the uncertainty of the sampled point decreases toward zero as $y_0$ approaches 0 and 1. This occurs when $\tau_0$ approaches $+\infty$ and 0 if there are no SPAM errors (i.e. $\alpha=\beta=0$), which are the limiting cases where we always measure the qubit in state $|0\rangle$ or $|1\rangle$. Inserting the relation between variation in $y_0$ and $\DecayRate$ we find
\begin{equation}
    \langle \delta\EstimateDecayRate \rangle = \frac{1}{(y_0-\beta) \tau_0 \sqrt{N}} \sqrt{y_0 - y_0^2},
\end{equation}
In the case of no SPAM errors, the uncertainty in $\EstimateDecayRate$ becomes:
\begin{equation}
    \langle \delta\EstimateDecayRate \rangle = \frac{1}{\tau_0 \sqrt{N}} \sqrt{y_0^{-1} - 1},
\end{equation}
and by substituting $y_0$, we get:
\begin{equation}
    \langle \delta\EstimateDecayRate \rangle = \frac{1}{\tau_0 \sqrt{N}} \sqrt{\exp(\DecayRate \tau_0) - 1}
    \label{eq:uncertainty_decay_rate}.
\end{equation}
This expression describes the evolution of the uncertainty in the exponential parameter as a function of the number of samples $N$ and the sampling time $\tau_0$. To minimize the uncertainty in $\DecayRate$, we solve:
\begin{equation}
    \frac{\mathrm{d}}{\mathrm{d}\tau} \langle \delta\EstimateDecayRate \rangle = 0.
\end{equation}
The optimal sampling point $\tau_0$ is found to be:
\begin{equation}
\label{tauop_vs_N}
    \tau_{\text{opt}} = \frac{\text{LambertW}(-2 e^{-2}) + 2}{\DecayRate},
\end{equation}
where the Lambert W function is the inverse relation of $f(w) = we^w$ as introduced in the previous section.  Approximately, the optimal value is:
\begin{equation}
    \tau_{\text{opt}} \approx 1.59 \DecayRate^{-1}.
\end{equation}
Using Eq.~(\ref{eq:uncertainty_decay_rate}), the minimal uncertainty in $\DecayRate$ is:
\begin{equation}
    \langle \delta\EstimateDecayRate \rangle_{\text{opt}} = \frac{\DecayRate}{1.59 \sqrt{N}} \sqrt{\exp(1.59) - 1} \approx \frac{1.24 \DecayRate}{\sqrt{N}}
\end{equation}
This result expresses the lowest achievable uncertainty as a function of the number of single-shot samples $N$. However, we recall that each measurement is inherently time-dependent, so it is more realistic to consider the time-dependent description of the previous subsection.

The analytical solution in \eqref{tauop_vs_N} can be also be extended to the specific case of SPAM errors where $\beta=0$, to get 
\begin{equation}
    \tau_{\text{opt}} = \frac{\text{LambertW}[2(\alpha-1) e^{-2}] + 2}{\DecayRate}.
\end{equation}
To our knowledge, there is no simple analytical solution for $\beta\neq0$.

\clearpage
\section{Decay-rate estimation with gamma distributions}
In this section we derive the moment matching of the gamma distribution $\Gamma(k,\theta)$ presented in the main text. Let the qubit decay rate $\lambda\sim{}\Gamma(k,\theta)$. Its probability density function (pdf) reads
\begin{equation}\label{eq:prior_Plambda}
    p(\lambda|k,\theta)=\frac{\theta^k}{\Gamma(k)}\lambda^{k-1}e^{-\theta\lambda}
\end{equation}
We make the prior assumption that $T_1=1/\lambda\sim{} \Gamma^{-1}(k,\theta)$, i.e., $T_1$ follows the inverse Gamma distribution. In the following, we will use $\text{E}[\cdot]$ to denote the expectation operator and refer to $X|Y$ as the random variable $X$ conditioned on $Y$. Together, $\text{E}[X|Y]$ is the expectation of $X$ after conditioning on $Y$.

Assuming the qubit is initialized to the (unobserved) excited state $s=1$, the probability that a qubit is still in the excited state after waiting time $\tau$ is
\begin{equation}
    P(s=1|\tau,\lambda)=e^{-\lambda \tau}\enspace.
\end{equation}
We can only observe the measured state $m$ which can differ from $s$ due to SPAM errors with error probabilities
$\alpha=P(m=0|s=1)$ and $\beta=P(m=1|s=0)$. Here, we assumed that the SPAM error parameters are independent of $T_1$ and $\tau$, as is commonly done in practice. Thus, we observe $m=1$ with probability
\begin{equation}\label{eq:prob_Pm}
    P(m=1|\tau,\lambda)=\beta P(s=0|\tau,\lambda) +(1-\alpha) P(s=1|\tau,\lambda)=\beta+(1-\beta-\alpha) e^{-\lambda \tau}\enspace.
\end{equation}
$P(m|\tau,\lambda)$ can be written in the form $a_m-b_me^{-\lambda \tau}$ with $a_m = (1 - m)(1 - \beta) + m \beta  $ and $b_m = (1 - 2m)(1 - \alpha - \beta)$.
After observing the value of the measurement $m$, the posterior distribution of $\lambda$ can be written as (using $Z$ as normalization constant of the posterior):
\begin{align}
\begin{split}\label{eq:exact_posterior}
    p(\lambda|k,\theta,m, \tau)&=\frac 1 Z p(\lambda|k,\theta)(a_m-b_me^{-\lambda \tau})\\
    &= \frac 1 Z\left( a_m p(\lambda|k,\theta)-b_m p(\lambda|k,\theta)e^{-\lambda \tau}\right)\\
    &=\frac { a_m p(\lambda|k,\theta)-b_m \left(\frac{\theta}{\theta+\tau}\right)^k p(\lambda|k,\theta+\tau)} {a_m -b_m\left(\frac{\theta}{\theta+\tau}\right)^k}
\end{split}
\end{align}
Note that if $a_m=0$ (which is the case for $m=1$ and $\beta=0$), the posterior simplifies to the gamma distribution $p(\lambda|k,\theta+\tau)$. However, in general we have that after $N$ measurements, the posterior is a linear combination of $2^N$ gamma distributions. In general, due to the law of large numbers, the posterior distribution will approach the normal distribution as $N\rightarrow + \infty$.

Since for $m=1$ at low SPAM error rate, the posterior is still approximately gamma distributed and the gamma distribution also approaches the normal distribution as $k\rightarrow + \infty$, our approach is to approximate the posterior $p(\lambda|k,\theta,m, \tau)$ using a gamma distribution via moment matching. Moment matching makes sense in this case since it can be implemented on the FPGA-powered controller but also enables the right scaling behavior of mean and variance, and thus ensures that as $N\rightarrow + \infty$, $k\rightarrow + \infty$.

The first and second moments of the gamma distribution are given by
\begin{equation}
    \text{E}[\lambda|k,\theta]=\frac k \theta,\;\text{E}[\lambda^2|k,\theta]=\frac {k+k^2}{\theta^2}\enspace.
\end{equation} 
Moment matching computes the parameters $k,\theta$ based on the moments. For the gamma distribution, we obtain
\begin{equation}
    \frac 1\theta = \frac{\text{E}[\lambda^2|k,\theta]}{\text{E}[\lambda|k,\theta]}-\text{E}[\lambda|k,\theta],\;k=\theta \text{E}[\lambda|k,\theta]
\end{equation}
Thus, by computing the moments of the posterior $p(\lambda|k,\theta,m, \tau)$ (using $a_m,b_m$ as above), we can compute the parameters of the approximating gamma distribution. Direct computation of the moment gives:
\begin{align*}
    \text{E}[\lambda|k,\theta,m,\tau]&=\frac { a_m\text{E}[\lambda|k,\theta]-b_m\left(\frac{\theta}{\theta+\tau}\right)^k \text{E}[\lambda|k,\theta+\tau]} {a_m-b_m\left(\frac{\theta}{\theta+\tau}\right)^k}\\
    &=\frac{k}{\theta}\frac {a_m-b_m\left(\frac{\theta}{\theta+\tau}\right)^{k+1}} {a_m-b_m\left(\frac{\theta}{\theta+\tau}\right)^k},
\end{align*}
\begin{align*}
    \text{E}[\lambda^2|k,\theta,m,\tau]&=\frac { a_m\text{E}[\lambda^2|k,\theta]-b_m\left(\frac{\theta}{\theta+\tau}\right)^k \text{E}[\lambda^2|k,\theta+\tau]} {a_m-b_m\left(\frac{\theta}{\theta+\tau}\right)^k}\\
    &=\frac {a_m\frac{k+k^2}{\theta^2}-b_m\left(\frac{\theta}{\theta+\tau}\right)^k \frac{k+k^2}{(\theta+\tau)^2}} {a_m-b_m\left(\frac{\theta}{\theta+\tau}\right)^k}\\
    &=\frac{k+k^2}{\theta^2} \frac {a_m-b_m\left(\frac{\theta}{\theta+\tau}\right)^{k+2} } {a_m-b_m\left(\frac{\theta}{\theta+\tau}\right)^k}\\
    &=\text{E}[\lambda|k,\theta,m, \tau]\frac{1+k}{\theta} \frac {a_m-b_m\left(\frac{\theta}{\theta+\tau}\right)^{k+2} } {a_m-b_m\left(\frac{\theta}{\theta+\tau}\right)^{k+1}}.
\end{align*}
Both equations are programmed on the controller for the adaptive estimation.
\clearpage
\section{Comparison of nonadaptive and adaptive Bayesian methods}
\begin{figure*}
\centering
\includegraphics{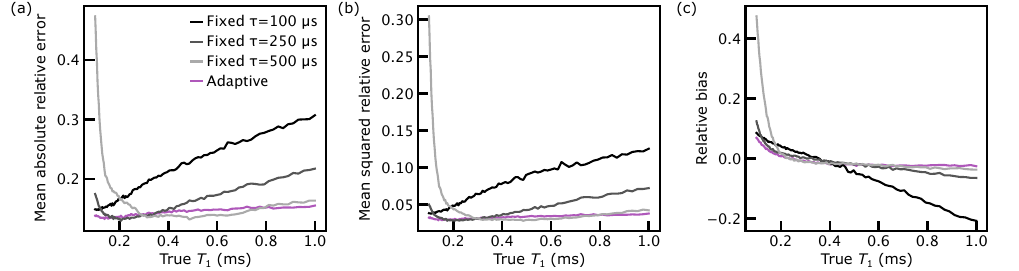}
\caption{\textbf{Comparison of nonadaptive and adaptive Bayesian methods}
For each simulated true $T_1$ value we show the
(a) mean absolute relative error $\langle| (T_1-\hat{T}_1)/T_1| \rangle$,
(b) mean squared relative error $\langle [ (T_1-\hat{T}_1)/T_1]^2 \rangle$, and
(c) relative bias $\langle\hat{T}_1/T_1\rangle-1$, where the mean is computed over 20,000 trials (see text for parameters used in the simulation).
}
\label{fig:FigS12}
\end{figure*}
In Fig.~\ref{fig:FigS12} we compare by simulation our adaptive method (purple lines) with a nonadaptive frequentist approach, where the waiting time $\tau$ is fixed (grayscale curves, with $\tau= \SI{100}{\micro\second}$, $\SI{250}{\micro\second}$ and $\SI{500}{\micro\second}$). We derive the frequentist approach as follows: We use the exact same probabilistic model with the same likelihood term \eqref{eq:prob_Pm} and prior \eqref{eq:prior_Plambda}. Using the sampled measurements $m_i\in \{0,1\}$ obtained with waiting time $\tau$, we then compute the maximum a-posteriori estimate
\[
    \hat{\lambda}= \operatorname*{arg\,max}_\lambda \left[ \log P(\lambda|k_0,\theta_0) + \sum_{i=1}^N \log P(m=m_i|\tau,\lambda) \right],
\]
and obtain the estimate $\hat{T}_1=1/\hat{\lambda}$. Note that the prior acts as a regularizer to the maximum log-likelihood estimate. This is necessary, since for low numbers of events $m_i=1$ or $m_i=0$ the maximum likelihood estimate of $\hat{\lambda}$ is either 0 or unbounded. Choosing the prior of our method as a regularizer allows for a fair comparison of the methods.

We plot the mean absolute relative error $\langle| (T_1-\hat{T}_1)/T_1| \rangle$ (a), mean squared relative error $\langle[ (T_1-\hat{T}_1)/T_1]^2 \rangle$ (b) and the relative bias $\langle\hat{T}_1/T_1\rangle - 1$ (c) as a function of the ``true'' $T_1$ over hundreds of microseconds spanned by our long-lived transmon qubits. For each true $T_1$ value, the mean is computed from 20,000 trials, where each trial consists of $N=100$ single-shot measurements. Similarly to the experiments in the main text, in the simulation we set $\alpha = \beta = 0.12$, and for the adaptive estimation each trial always starts with the same gamma prior distribution with $(k_0,\, \theta_0 ) = (3,\,\SI{450}{\micro\second})$. The simulation does not take into account finite idle times $t$ in the setup, thus we simply set $c=1$ for the adaptive method.

The results show three points: (i) The adaptive choice of $\tau$ makes the algorithm robust with approximately constant low error rates (a,b), across the whole range of interest (hundreds of microseconds), while the frequentist method with fixed $\tau$ achieves similarly low errors only when $\tau\approx T_1$. (ii) Our method exhibits very low bias in the region $T_1>\SI{150}{\micro\second}$ and the lowest absolute bias overall. For the region $T_1<\SI{150}{\micro\second}$ bias is introduced via the choice of prior which only has little probability mass in this region. (iii)  The approximated inference scheme using moment matching works very well and there is only little information lost.

\section{How SPAM affects the estimation protocol}

\begin{figure*}
\centering
\includegraphics{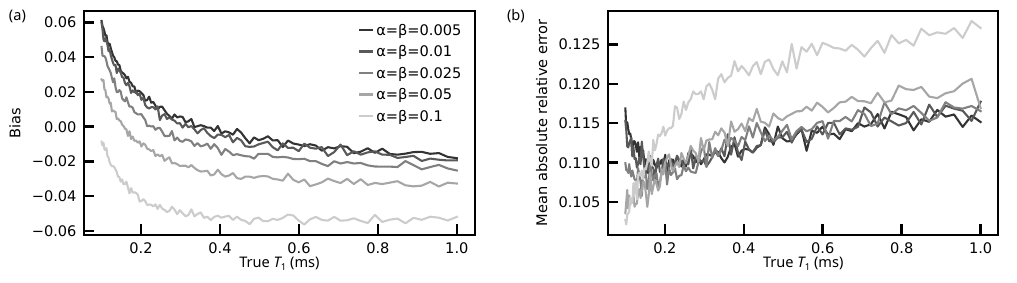}
\caption{\textbf{How SPAM affects the estimation protocol.}
For each simulated true $T_1$ value we show  
(a) the relative bias $\langle \hat{T}_1/T_1\rangle - 1$ and  
(b) the mean absolute relative error $\langle |(T_1-\hat{T}_1)/T_1| \rangle$,  
computed over 10{,}000 trials and different values of $\alpha=\beta$.  
The true SPAM parameters are fixed to $\alpha_{\mathrm{sim}}=\beta_{\mathrm{sim}}=0.025$, while the estimator uses mismatched values $\alpha_{\mathrm{est}}=\beta_{\mathrm{est}}$.}
\label{fig:FigS13}
\end{figure*}

To assess the influence of SPAM miscalibration on the adaptive Bayesian estimator, we fix the true SPAM parameters to $\alpha_{\mathrm{sim}}=\beta_{\mathrm{sim}}=0.025$ and sweep the estimator values $\alpha=\beta$. The resulting bias and mean absolute relative error are shown in Fig.~\ref{fig:FigS13}. Overestimating SPAM errors (larger $\alpha,\,\beta$, lighter curves) consistently worsens performance across all $T_1$. This is consistent with the estimator treating measurements as less informative, which slows the posterior update after each single-shot measurement and yields larger residual bias and variance after the fixed number of samples. Underestimating SPAM errors (smaller $\alpha,\,\beta$, darker curves) leads to increased error primarily at small $T_1$. In this regime the estimator overweights observed ground-state outcomes, effectively overestimating $T_1$ and producing a positive bias visible in Fig.~\ref{fig:FigS13}(a). Using the correct SPAM parameters ($\alpha_{\mathrm{est}}=\beta_{\mathrm{est}}=0.025$) yields the smallest bias and lowest overall error [panel (b)].

\section{Validity of the gamma approximation}

\begin{figure*}
\centering
\includegraphics{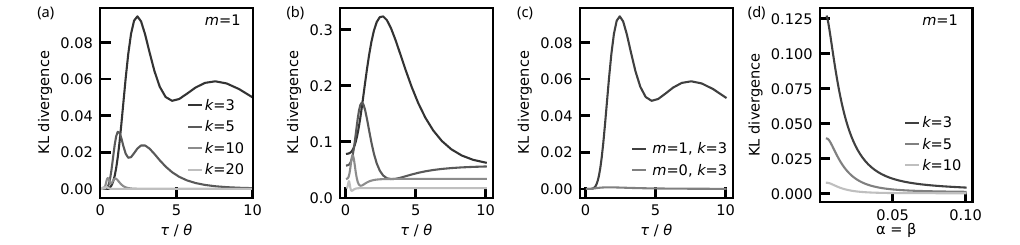}
\caption{\textbf{Comparison of exact posteriors with gamma and truncated-normal approximations.} For $\alpha=\beta = 0.01$, 
(a) KL divergence between the exact posterior and the gamma approximation as a function of $\tau/\theta$ for several prior shape parameters $k$. The maximal deviation occurs for outcome $m=1$  and decreases rapidly with increasing $k$.
(b) Corresponding KL divergence for the truncated-normal approximation.
(c) KL divergence for $k=3$ comparing outcomes $m=0$ and $m=1$, showing that the gamma approximation
is worse in the skewed $m=1$ case, while updates with $m=0$ remain very close to the gamma family.
(d) Dependence of the KL divergence on the SPAM error probability, $\alpha=\beta$, evaluated at the worst-case $\tau$ for each $k$. Larger error rates reduce the measurement information and thus keep the posterior closer to the prior, improving the gamma approximation.}
\label{fig:FigS14}
\end{figure*}

\paragraph{KL divergence between exact and approximate posteriors}

The posterior after a single measurement depends only on the ratio $\tau/\theta$ of the waiting time to the prior scale parameter. We therefore fix $\theta=1$ without loss of generality~
\footnote{This will not influence the quantitative outcome of the integral, since it merely rescales the integration axis. The key point is that a change of $\theta$ only leads to a scaling of $\lambda$, i.e., it is a linear transformation of the $\theta=1$ case. This holds for both the exact distribution and our moment matching. It then follows that the KL divergence is invariant under affine transformations of the compared random variable.} and compute the Kullback--Leibler (KL) divergence 
\begin{equation}
D_{\mathrm{KL}}(P_{\mathrm{exact}} \,\|\, Q)
=
\int_{0}^{\infty}
p_{\mathrm{exact}}(\lambda)\,
\log\!\left[
\frac{
p_{\mathrm{exact}}(\lambda)
}{
q(\lambda)
}
\right]
\, \mathrm{d}\lambda,
\end{equation}
where $P_{\mathrm{exact}}$ is the exact posterior [Eq.~(\ref{eq:exact_posterior})] $Q$ is either the gamma approximation or the truncated normal approximation. The exact posterior can be evaluated in closed form as a linear combination of two gamma densities, whereas the approximations are obtained by matching the first two posterior moments.

Figure~\ref{fig:FigS14}(a) shows the KL divergence between
the exact posterior and the gamma approximation for the worst-case outcome $m=1$, where deviations are largest, using $\alpha=\beta = 0.01$. We plot the divergence as a function of $\tau/\theta$ for several prior shape parameters $k\in\{3,5,10,20\}$. Two trends are apparent. First, the KL divergence is bounded and remains small ($\lesssim 0.1$) over the entire parameter range. Second, the approximation improves rapidly with increasing $k$, as expected because the posterior becomes sharply peaked and hence more Gaussian-like.

\paragraph{Comparison with a truncated normal approximation} To verify whether the gamma choice is better with respect to a Gaussian distribution, we compare our moment-matched gamma approximation with a moment-matched truncated normal distribution. As shown in Fig.~\ref{fig:FigS14}(b), the truncated normal typically exhibits a much larger KL divergence than the gamma function in (a), especially for small $k$ where the posterior is skewed. This confirms
that the gamma family captures the asymmetry of the posterior more faithfully than a normal distribution. Only for large $k$ the two approximations become comparable, which is consistent with the central limit theorem. 

Figure~\ref{fig:FigS14}(c) compares the cases
$m=0$ and $m=1$ for fixed $k=3$. As expected, the approximation error is largest for $m=1$, which shifts the posterior toward larger decay rates and produces a more asymmetric shape. In contrast, the $m=0$ update preserves the overall shape of the prior and is therefore very well captured by a gamma distribution. We notice that for $\alpha=0$, the gray line ($m=0$, $k=3$) is 0.

\paragraph{Dependence on SPAM} Here we investigate how SPAM influences the quality of the approximation. To this end, we sweep $\alpha=\beta$ from $0.005$ to $0.1$, and evaluate the KL divergence at the value of $\tau$ that produces the worst-case approximation error for each $k$ [identified from Fig.~\ref{fig:FigS14}(a)]. As shown in Fig.~\ref{fig:FigS14}(d), the KL divergence decreases monotonically with increasing error rates. This behavior is intuitively expected: for large $\alpha$ and $\beta$, the measurement contains little information and the posterior remains close to the prior, which is exactly gamma-distributed. Consequently, the gamma approximation is most challenged in the experimentally relevant regime of small but nonzero infidelity, yet even in this case the divergence remains below $\approx 0.1$.

In summary, the KL-divergence analysis supports three conclusions: (i) the gamma approximation provides an accurate representation of the exact posterior over a wide parameter range, with errors on the level of a few percent in KL divergence, (ii) it consistently outperforms a truncated normal approximation, particularly in the skewed, low-$k$ regime, and (iii) the approximation improves both with increasing prior information (larger $k$) and with increasing measurement error rates, where the posterior approaches the gamma-distributed prior.

\clearpage

\section{Quasistatic approximation and Markovianity}

In this section we comment on the quasistatic approximation mentioned in the main text. We start by invoking standard derivation of Fermi Golden Rule, which amounts to a linear approximation of the exponential decay. Within this approximation the probability of occupying the ground state after the time $t$ is given by:
\begin{equation}
P_0(t) \approx  \frac{4|M_{01}|^2}{\hbar^2} \int_{-\infty}^{\infty} d\omega S_V(\omega) \frac{\sin^2((\omega - \omega_{01})t/2)}{(\omega - \omega_{01})^2}
\end{equation}
where $S_V(\omega)$ is the spectral density of the noise that couples to the qubit, and $\omega_{01}$ is the qubit frequency and $|M_{01}|^2$ is the matrix element for the transition. In this context, \textit{Markovian} approximation is typically associated with the fact that the noise spectrum $S_V(\omega)$ is approximately constant over the bandwidth $\sim 1/t$ around $\omega_{01}$. In this case the integral can be evaluated, yielding a constant rate $P_0(t) \approx \DecayRate t$, where $\DecayRate = 2\pi |M_{01}|^2 S_V(\omega_{01})/\hbar^2$. 

This shows that a time-dependent rate, $\DecayRate(t)$, can arise in three ways. First, the matrix element $|M_{01}(t)|^2$ is time-dependent, which does not affect the Markovian nature of the decay. Alternatively the power of the noise can be time-dependent due to either (i) change of the spectrum $S_V(\omega, t)$ or (ii) change of the qubit frequency $\omega_{01}(t)$. In both cases the decay remains Markovian as long as the noise is broadband on the scale $\sim 1/t$.

When $\DecayRate(t)$ varies during the measurement, the master equation $\frac{d}{dt} P_0(t) = -\DecayRate(t) P_0(t)$ has the solution:
\begin{equation}
    P_0(t) = \exp\left(-\int_0^t \DecayRate(t') dt'\right) = \exp\left(- \overline{\DecayRate} t \right)
\end{equation}
where $\overline{\DecayRate} = \frac{1}{t}\int_0^t \DecayRate(t') dt'$ is the time-averaged decay rate over the measurement window $t$.

It shows that our Bayesian estimation, which fits an exponential decay to find a single rate, is measuring the time-averaged rate, $\overline{\DecayRate}$, correctly capturing the mean decay effect even if the rate fluctuates rapidly during the measurement. Moreover, since $\DecayRate(t)$ is a stochastic process, $\overline{\DecayRate}(t)$ is a random variable. But this is also how Bayesian statistics treats an estimated variable, and so the method should be robust to whatever fast fluctuations are not resolved (which should just show up as some inherent uncertainty in the $\DecayRate$ the controller is trying to estimate).

This allows us to clarify what our scheme estimates across different fluctuation timescales:
\begin{itemize}
    \item \textbf{Slow Fluctuations} (slower than our total estimation time): Our scheme can successfully \emph{track} these slow drifts in the average decay rate in real-time. This is the primary demonstration in our manuscript.
    \item \textbf{Fast Fluctuations} (much faster than the single measurement time $t$): Our scheme will yield a stable estimate of the mean decay rate $\langle \DecayRate \rangle$. The fast fluctuations are effectively averaged out during each measurement.
    \item \textbf{Intermediate Fluctuations} (timescale comparable to $t$): For fluctuations in this regime, each estimation shot witnesses a slightly different time-average. Presence of such fluctuations, might be responsible for generating uncorrelated outliers, if the prior distribution becomes too narrow. However, as shown empirically, such events are sufficiently rare.
\end{itemize}

\clearpage

\section{Theoretical estimation uncertainty limit}
\begin{figure*}
\centering
\includegraphics{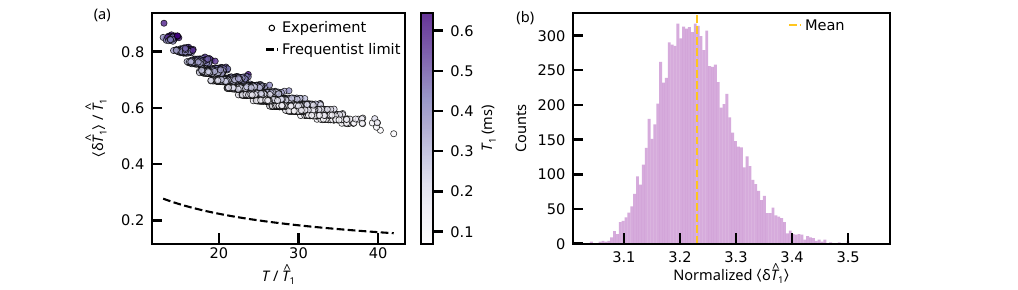}
\caption{\textbf{Theoretical estimation uncertainty limit.}
(a) $N=10,000$ experimental uncertainties $\langle \delta \hat{T}_1 \rangle$ (68\% credible interval) as a function of total time budget $T$, both axes normalized by the corresponding $\hat{T}_1$. The dashed line corresponds to the frequentist limit given by Eq.~(\ref{eq:frequentist_limit}). (b)  $N=10,000$ experimental uncertainties normalized by the frequentistic limit, the average is represented by the dashed yellow line.}
\label{fig:FigS9}
\end{figure*}
In this section we test the performance of our Bayesian protocol compared to the ideal case in the absence of SPAM ($\alpha = \beta = 0$). Then the expected uncertainty from the estimation $\langle \DecayRate \rangle$ from Eq.~(\ref{eq:uncertainty_gamma_finite_time}) becomes:
\begin{equation}
    \langle \delta\EstimateDecayRate \rangle = \sqrt{\frac{1-e^{-\DecayRate \tau}}{\tau Te^{-\DecayRate \tau}}}
\end{equation}
In the limit of negligible idle times in the setup, $t\rightarrow0$ and thus $\tau_{\text{opt}}\rightarrow0$ [cf. Fig.~\ref{fig:FigS4}(b)]. Then one obtains what we refer to as the ``frequentist limit":
\begin{equation}
    \langle \delta\EstimateDecayRate \rangle \approx \sqrt{\frac{\DecayRate}{T}},
\end{equation}
which is equivalent to 
\begin{equation}
    \langle \delta\hat{T}_1 \rangle \approx T_1 \sqrt{\frac{T_1}{T}}.
    \label{eq:frequentist_limit}
\end{equation}

To compare the frequentist limit with the experimental Bayesian estimate, we perform $10,000$ repetitions of adaptive $\DecayRate$ estimations, using $30$ single-shot measurements per repetition. The settings are the same as in Fig.~1(d) of the main text.

In Fig.~\ref{fig:FigS9}(a), we plot the uncertainties $\langle \delta \hat{T}_1 \rangle$ obtained from $10,000$ repetitions, each requiring an elapsed laboratory time of $T$ of a few milliseconds. Both $\langle \delta \hat{T}_1 \rangle$ and $T$ are normalized by their respective estimates $\hat{T}_1$. The measured uncertainty scales as predicted by Eq.~(\ref{eq:frequentist_limit}) (dashed line).

In Fig.~\ref{fig:FigS9}(b), we present a histogram of the experimental uncertainties $\langle \delta \hat{T}_1 \rangle$ (68\% credible interval), normalized by their corresponding frequentist limit $\langle \delta\hat{T}_1 \rangle$. On average, the Bayesian protocol yields an uncertainty approximately 3.23 times higher than the limit set by Eq.~(\ref{eq:frequentist_limit}). We emphasize that the frequentist limit applies to an unbiased estimator, whereas in our case, the experimental uncertainty $\langle \delta \hat{T}_1 \rangle$ is strongly determined by the initial choice of the prior distribution. 

\clearpage
\section{Validation of Estimated values of $T_1$}
\begin{figure*}
\centering
\includegraphics{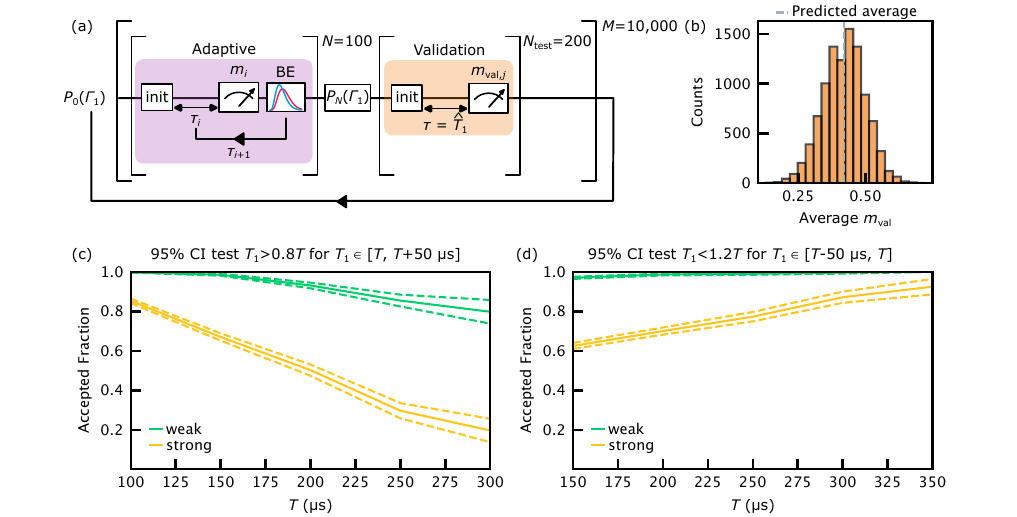}
\caption{\textbf{Validation of estimated values of $T_1$.}
(a) One loop (solid arrow) represents a single repetition of the validation protocol. In this experiment, the controller performs $ M = 10,000 $ repetitions. At the beginning of each repetition, the controller performs $ N=100 $ adaptive probing cycles (Adaptive). To test whether the estimated $ T_1 $ is correct, the estimation sequence is followed by $ N_{\text{test}} = 200 $ cycles (Validation), where the qubit is initialized to the excited state and the same waiting time $ \tau = \hat{T}_1 $ is used before measurement ($ m_{\text{lin},i} $).  
(b) Histogram of the number of counts of the average $m_\text{val}$ extracted from each of the $10,000$ validation sequences. The dashed line is the predicted average of the distribution based on the choice of $ \tau = \hat{T}_1 $.
(c-d) Weak and strong tests (see text) to validate b: $T_1>0.8\, \hat{T}_1$, d: $T_1<1.2\, \hat{T}_1$.
}
		\label{fig:FigS5}
	\end{figure*}
    
With a small number of measurements, we cannot obtain a sufficiently high precision estimate of $T_1$. Thus, for practical applications, it is most important that if the algorithm returns a large value for the mean of the posterior $\hat{T}_1$, the real value of $T_1$ is with high probability large enough to allow for higher-fidelity qubit operations. We therefore implemented statistical tests to validate that $T_1>0.8\, \hat{T}_1$ or, for the opposite direction, $T_1<1.2\, \hat{T}_1$, each indicating at most a 20\% over-or underestimate. 

We implement this as follows for the test of $T_1>0.8\, \hat{T}_1$, the other direction follows analogously. Let us assume the value of $T_1$ at the end of the runtime of the algorithm and during the test measurements is $T_1=(1+q)\hat{T}_1$, for some $q\in (-1,1)$. Then in our experiment with waiting time $\hat{T}_1$ and known SPAM error rates $\alpha$ and $\beta$ the probability to observe state $m=1$ is 
\[
    P(m=1|q)=\beta+ (1-\alpha-\beta) \exp\left(-\frac{1}{1+q}\right)
\]
Under this assumption, the number of measurements with $m=1$ among $N_{\text{test}}$ measurements follows a binomial distribution, $S\sim{}\text{Binomial}(N_{\text{test}},P(m=1|q))$. If $T_1>(1+q)\hat{T}_1$ we would expect to observe larger values of $S$ on average (and smaller values for $T_1<(1+q)\hat{T}_1$). 

Based on this, we devised two statistical tests, which we call the ``weak" and ``strong" test. Assume we want to know whether  $T_1>(1+q)\hat{T}_1$ for some chosen value of $q$. For the weak test, we create $95\%$ confidence interval $[S_\text{weak},N_{\text{test}}]$ of the binomial distribution, i.e.
$P(S\in [S_\text{weak},N_{\text{test}}])\geq0.95$. If we observe $S<S_\text{weak}$, then the data provides evidence against the hypothesis that $T_1>(1+q)\hat{T}_1$ and thus we conclude that the algorithm likely returned a false result. However, due to a limited number of validation measurements $N_{\text{test}}$ we can take before device drift affects the results, and the presence of SPAM errors, we might not be able to reject many wrong results. For this reason, we then devised a second test, using the confidence interval $P(S\in [0,S_\text{strong}])\geq0.95$. With this, we accept the hypothesis only when $S>S_\text{strong}$. This test is equivalent to rejecting the counter hypothesis $T_1<(1+q)\hat{T}_1$ and thus the data provides evidence that $T_1$ is at least as large as tested. This test will tend to reject correct estimation results with small $N_{\text{test}}$.

To implement this framework, we perform the following experiment. For each repetition [see Fig.~\ref{fig:FigS5}(a)], the controller executes $N = 100$ probing cycles to estimate $\DecayRate$, using a gamma prior distribution with parameters $(k, \theta) = (3,\,\SI{450}{\micro\second})$. After the estimation sequence, a validation sequence follows. During the validation, the controller performs $N_{\text{test}} = 200$ qubit cycles, initializing the qubit in the excited state and setting the waiting time to the estimated $\hat{T}_1$ from the previous adaptive probing sequence. After waiting for $\tau \approx \hat{T}_1$, the qubit state is measured and classified as $m_{\text{val}, j}$. 

The interleaved adaptive and validation sequences are repeated $M = 10,000$ times. In this experiment, we set $\alpha = 0.105$ and $\beta = 0.14$. Figure~\ref{fig:FigS5}(b) shows a histogram obtained by binning the average $\langle m_{\text{val}} \rangle$ over $j = 1,2,\dots,200$ for each of the 10,000 validation sequences. The distribution is approximately centered around the expected probability $p=\beta+(1-\alpha-\beta)e^{-1} \approx 0.418$ of measuring the excited state after setting $\tau \approx 1/\DecayRate$ [dashed line in Fig.~\ref{fig:FigS5}(b)].

For the test, we then partition the data and select the experiments for which $\hat{T}_1$ falls in an interval $[T,T+\SI{50}{\micro\second}]$ for $T\in \{100,150,200,250,300\} \SI{}{\micro\second}$ (the proportion of estimation results outside this interval is too small to evaluate). For all instances of the experiment in such interval, we compute the weak and strong tests validating the hypotheses $T_1>0.8\,T$ and $T_1<1.2\,T$ and record the fraction of passed tests as well as their 95\% confidence interval based on the normal distribution.

The results [see Fig.~\ref{fig:FigS5}(c-d)] show that for both experiments the weak test succeeded in between 80\% to 100\% of the estimation results in each interval. Thus, for the vast majority of the results the estimation errors are not so large that $N_{\text{test}}=200$ samples can reject the hypothesis that the estimation error is below 20\%. However, for the strong test, we see a much lower acceptance rate, i.e., the estimation errors are also not small enough to reject the opposite hypothesis that the error is larger than 20\%. We can see in both cases that the fraction of accepted results increases towards more extreme results of $\hat{T}_1$. Thus, Fig.~\ref{fig:FigS5}(c) indicates that with increasing value of estimated $\hat{T}_1$, the estimation error increases, as expected from Eq.~(\ref{eq:frequentist_limit}). For those intervals with large $T$, the fraction of successful tests Fig.~\ref{fig:FigS5}(d) is also increasing. Both tests indicate that large values of $\hat{T}_1$ might be overestimates of $T_1$. The opposite holds for the other extreme for $T\leq \SI{200}{\micro\second}$, thus small values of $\hat{T}_1$ might be underestimates.

However, these results do not take into account the possible drift of $T_1$ during the validation measurements. Especially for large $\hat{T}_1$, taking $N_{\text{test}}=200$ validation samples can take a long time. For example for $\hat{T}_1=\SI{250}{\micro\second}$, the time taken to estimate a validation sample is $\SI{55}{\milli\second}$ in addition to the algorithm runtime. Thus, device drift may have already changed the results. This might be consistent with the smaller number of rejected samples with small $T_1$ which can take the samples in approximately half the time.
\clearpage
\section{Frequency of events of large switches in $T_1$}
In this section we describe the analysis mentioned in the Section IV.C (Power spectral density and Allan deviation) of the main text where we conclude that switches in $T_1$ greater than $\SI{100}{\micro\second}$ on average occur once every $\SI{7.7}{\second}$.

To better understand the estimation errors of our algorithm and the fine-grained structure of the noise on $T_1$, we partition all the single-shot outcomes of the 3-day-long time trace of Fig.3(a) of the main text into intervals of maximum length $\SI{200}{\milli\second}$. We ensure that partitioning only occurs after the controller completes an estimation repetition, as the single-shot outcomes within a repetition are not statistically independent due to their waiting times depending on previous outcomes. Since each estimation repetition is initialized with a prior independent of previous results, the repetition estimates are statistically independent. Within each interval, we split the estimates into a training and test set by assigning every second repetition run to the training set (e.g., runs 1, 3, 5, $\dots$) and the remainder to the test set (e.g., 2, 4, 6, $\dots$). We average the $T_1$ estimates from the training set to obtain an interval-wise estimate of the mean $\overline{T}_1$.

Next, we identify pairs of consecutive intervals that satisfy the following criteria: (a) the average $\overline{T}_1$ in both intervals lies within the range $\SI{100}{\micro\second} < \overline{T}_1 < \SI{400}{\micro\second}$, and (b) the difference between the average $\overline{T}_1$ values of the two intervals exceeds $\SI{100}{\micro\second}$. These thresholds are based on an independent validation, showing that individual estimates in this range have an error below 20\% with high probability. This filtering step retains 3.8\% of all $\SI{200}{\milli\second}$ intervals.

We then use the corresponding test sets to evaluate whether the difference in $\hat{T}_1$ is statistically significant. Let $\hat{T}_1^{L}$ and $\hat{T}_1^{R}$ denote the $\hat{T}_1$ estimates of the left and right intervals, respectively, and $T_1^{L}$, $T_1^{R}$ their unknown true values. Assuming without loss of generality that $\hat{T}_1^{L} < \hat{T}_1^{R}$, we perform two one-sided hypothesis tests: one to reject $H_0: T_1^{L} > \hat{T}_1^{R}$ using the left interval's test data, and one to reject $H_0: T_1^{R} < \hat{T}_1^{L}$ using the right interval's test data. Each test is performed at a 97.5\% confidence level, and a switch is considered verified when both tests succeed.

Out of the 3.8\% of intervals that pass the initial filter, 69\% yield a verified switch, corresponding to approximately 2.6\% of all intervals. This translates to one $T_1$ switch (exceeding $\SI{100}{\micro\second}$) every $\SI{7.7}{\second}$  on average. Note that the analysis is constrained by the limited sampling rate, especially for large $\hat{T}_1$ values, as the algorithm's runtime scales approximately linearly with $\hat{T}_1$ due to longer waiting times.

\clearpage
\section{Fit to Allan Deviation and Power Spectral Density}
\begin{figure*}
\centering
\includegraphics[width=\textwidth]{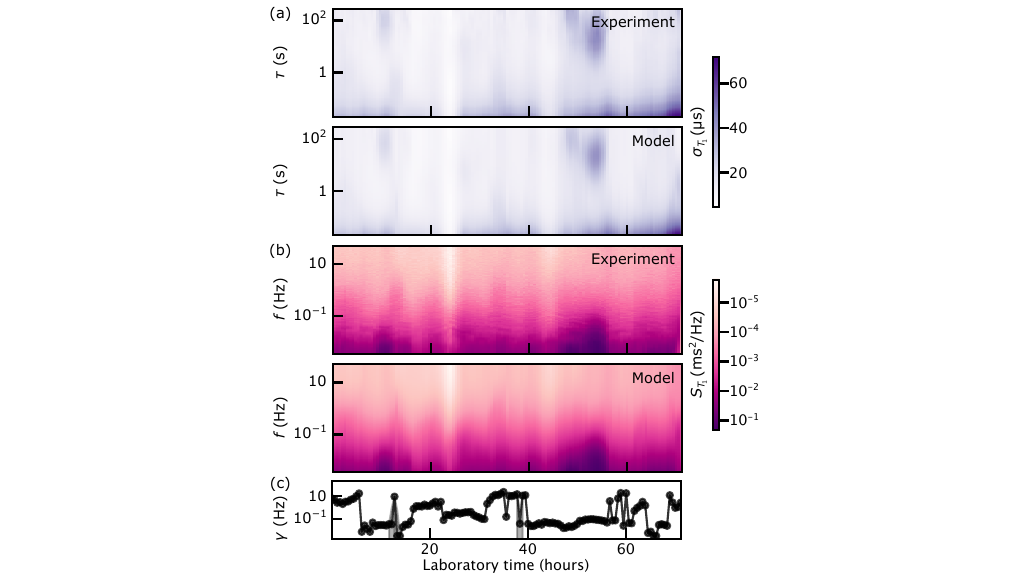}
\caption{\textbf{Comparing Allan deviation and PSD of $T_1$ time series with fit.}
(a) Extracted and fitted Allan deviation for the 72-hour-long estimates using 2.7-hour-long time intervals. (b) Extracted and fitted PSD for the 72-hour-long estimates using 2.7-hour-long windows with 80\% overlap.
(c) Switching rates $\gamma$ of the Lorentzian model used to fit the experimental data in panels (a,\,b).
}
\label{fig:FigS7}
	\end{figure*}
In this section we provide more details on the fit used for Fig.~3 and 4 of the main text.
We treat the estimated values of $\hat{T}_1$ as a stationary time series data and analyze the noise correlation using the power spectral density (PSD), defined as:  
\begin{equation}
	S_{\hat{T}_1}(f) = \int_{-\infty}^{+\infty} R_{\hat{T}_1}(\tau) e^{-2 \pi i  f \tau} \, \text{d}\tau,
\end{equation}
where $ R_{\hat{T}_1}(\tau) = \langle \hat{T}_1(\tau) \hat{T}_1(0)\rangle $ is the autocorrelation function of the estimated $\hat{T}_1$ values, and $\langle \ldots \rangle$ denotes the average over the sequence of estimated $\hat{T}_1$ values. To estimate the PSD, we use the Welch method, which divides the time series data into overlapping segments and computes the PSD for each segment. The PSD is then averaged over all segments to obtain the final estimate. The spectral density obtained for the 72-hour dataset is shown in Fig.~3(b) of the main text.  

As discussed in the main text, the obtained PSD is dominated by $1/f$ noise, with a spectrum $ S_{T_1}(f) = A_1/f $, suggesting the presence of low-frequency drift in the relaxation time. However, such a $1/f$-dominated PSD does not provide meaningful insight into other noise sources. For this reason, we also compute the Allan deviation, a method for measuring frequency stability, defined as:  
\begin{equation}
	\sigma_{\hat{T}_1}(\tau) = \sqrt{\frac{1}{2\tau^2} \bigg\langle \Big(\hat{T}_1[(n+2)\tau] - 2\hat{T}_1[(n+1)\tau] + \hat{T}_1[n\tau]\Big)^2 \bigg\rangle},
\end{equation}
where $ \hat{T}_1[n\tau] $ is the estimated $\hat{T}_1$ at the $n$-th time point. The Allan deviation is shown in Fig.~3(c) of the main text. In Fig.~\ref{fig:FigS7}, we plot the comparison between the measured and fitted Allan deviation [panel (a)] and PSD [panel (b)], as well as the fitted switching rate $\gamma$ [panel~(c)] for the Lorentzian process. The visual comparison of the experiment and model, along with the small uncertainty in $\gamma$, show that the fit captures the main features of the fluctuations in the estimated $\hat{T}_1$ values.

\bibliography{suppl_my_bibliography}